\documentclass[pra,aps,superscriptaddress,floatfix,nofootinbib,flushbottom,twocolumn]{revtex4-1}

\usepackage{graphics}
\usepackage{graphicx}
\usepackage{epsfig}
\usepackage{amsmath}
\usepackage{color}
\usepackage{xfrac}
\usepackage{cancel}
\usepackage{physics}
\usepackage[load=addn]{siunitx}
\usepackage{etoolbox}
\usepackage{multirow}

\DeclareSIUnit\ecm{\text{\em{e}}\cdot\cm}
\DeclareSIUnit\flux{\text{atoms}\per\second}
\DeclareSIUnit\torr{\text{Torr}}
\DeclareSIUnit\ralw{\gamma_{\text{Ra}}}

\makeatletter
\appto\abstract{%
  \let\latexlist\list%
  \def\list{\edef\keeprightskip{\the\rightskip}\latexlist}%
  \patchcmd\latexlist{\ignorespaces}{\rightskip\keeprightskip\ignorespaces}{}{}%
}
\makeatother

\newcommand{\Ra}{$^{225}$Ra}
\newcommand{\Sr}{$^{88}$Sr}
\newcommand{\iso}[2]{$^{#2}$#1}

\begin{document}

\title{Characterizing the Optical Trapping of Rare Isotopes by Monte Carlo Simulation}

\author{D.~H.~Potterveld}
\affiliation{Physics Division, Argonne National Laboratory, Argonne, Illinois 60439, USA}
\author{S.~A.~Fromm}
\email{fromm@nscl.msu.edu}
\affiliation{National Superconducting Cyclotron Laboratory and Department of Physics and Astronomy, Michigan State University, East Lansing, Michigan 48824, USA}
\author{K.~G.~Bailey}
\affiliation{Physics Division, Argonne National Laboratory, Argonne, Illinois 60439, USA}
\author{M.~Bishof}
\affiliation{Physics Division, Argonne National Laboratory, Argonne, Illinois 60439, USA}
\author{D.W. Booth}
\affiliation{Physics Division, Argonne National Laboratory, Argonne, Illinois 60439, USA}
\author{M.~R.~Dietrich}
\affiliation{Physics Division, Argonne National Laboratory, Argonne, Illinois 60439, USA}
\author{J.~P.~Greene}
\affiliation{Physics Division, Argonne National Laboratory, Argonne, Illinois 60439, USA}
\author{R.~J.~Holt}
\affiliation{Physics Division, Argonne National Laboratory, Argonne, Illinois 60439, USA}
\affiliation{Kellogg Radiation Laboratory, California Institute of Technology, Pasadena, California 91125, USA}
\author{M.~R.~Kalita}
\altaffiliation[Present Address: ]{TRIUMF, 4004 Wesbrook Mall, Vancouver, British Columbia V6T 2A3, Canada}
\affiliation{Physics Division, Argonne National Laboratory, Argonne, Illinois 60439, USA}
\affiliation{Department of Physics and Astronomy, University of Kentucky, Lexington, Kentucky 40506, USA}
\author{W.~Korsch}
\affiliation{Department of Physics and Astronomy, University of Kentucky, Lexington, Kentucky 40506, USA}
\author{N.~D.~Lemke}
\affiliation{Department of Physics and Engineering, Bethel University, St. Paul, MN 55112, USA}
\author{P.~Mueller}
\affiliation{Physics Division, Argonne National Laboratory, Argonne, Illinois 60439, USA}
\author{T.~P.~O'Connor}
\affiliation{Physics Division, Argonne National Laboratory, Argonne, Illinois 60439, USA}
\author{R.~H.~Parker}
\altaffiliation[Present Address: ]{Department of Physics, University of California at Berkeley, Berkeley, California 94720, USA}
\affiliation{Physics Division, Argonne National Laboratory, Argonne, Illinois 60439, USA}
\author{T.~Rabga}
\affiliation{Physics Division, Argonne National Laboratory, Argonne, Illinois 60439, USA}
\affiliation{National Superconducting Cyclotron Laboratory and Department of Physics and Astronomy, Michigan State University, East Lansing, Michigan 48824, USA}
\author{J.~T.~Singh}
\affiliation{National Superconducting Cyclotron Laboratory and Department of Physics and Astronomy, Michigan State University, East Lansing, Michigan 48824, USA}

\begin{abstract}
Optical trapping techniques are an efficient way to probe limited quantities 
of rare isotopes.  In order to achieve the highest possible measurement 
precision, it is critical to optimize the optical trapping efficiency.  This 
work presents the development of a three-dimensional semi-classical Monte 
Carlo simulation of the optical trapping process and its application to 
optimizing the optical trapping efficiency of Radium for use in the search of 
the permanent electric dipole moment of \Ra.  The simulation includes an 
effusive-oven atomic beam source, transverse cooling and Zeeman slowing of an 
atomic beam, a three-dimensional magneto-optical trap, and additional 
processes such as collisions with residual gas molecules.  We benchmark the 
simulation against a well-characterized \Sr\ optical trap before applying it 
to the \Ra\ optical trap.  The simulation reproduces the relative gains in 
optical trapping efficiency measured in both the \Sr~and \Ra~optical traps.  
The measured and simulated values of the overall optical trapping efficiencies 
for \Sr\ are in agreement; however, they differ by a factor of $30$ for \Ra.  
Studies of several potential imperfections in the apparatus or systematic 
effects, such as atomic beam source misalignment and laser frequency noise, 
show only limited effects on the simulated trapping efficiency for \Ra.  We 
rule out any one systematic effect as the sole cause of the discrepancy 
between the simulated and measured \Ra~optical trapping efficiencies; but, we 
do expect that a combination of systematic effects contribute to this 
discrepancy.  The accurate relative gains predicted by the simulation prove 
that it is useful for testing planned upgrades to the apparatus.
\end{abstract}

\date{\today}
\maketitle

\section{Introduction}\label{sec:intro}

Optical trapping is now a standard technique in a variety of rare isotope 
applications such as searches for new physics.  Examples include ground water 
dating with \iso{Kr}{81} and \iso{Ar}{39}~\cite{grl04,ar39prl}, precision 
measurements of the charge radii of exotic \iso{He}{6,8} nuclei~\cite{Lu2013,
Mueller2007,Wang2004}, a search for new physics by probing $\beta$-decay in 
unstable \iso{He}{6}, \iso{Na}{21}, \iso{K}{37}~\cite{na21prc,k37prl,ppnp19}, 
and a search for atomic parity violation in Fr isotopes~\cite{Tandecki2006}.  
Our primary application is using optically trapped \Ra~in the search for a 
permanent electric dipole moment (EDM)~\cite{Parker2012} (for a recent review 
of EDMs see~\cite{rmp2019}).  \Ra~offers enhanced sensitivity in diamagnetic 
atomic EDM searches due to the octupole deformation of its nucleus~\cite
{Auerbach1996,Dobaczewski2005}.  Using optically trapped \Ra~atoms, we have 
measured the \Ra~EDM to be $d\qty(\text{\Ra})<\SI{1.4e-23}{\ecm}$~\cite
{Bishof2016}.  In order to efficiently utilize limited quantities of \Ra~($10^
{14}$ atoms per measurement) and to further improve the measurement 
statistical precision, it is essential to optimize the optical trapping 
efficiency of our experimental apparatus.  Since it is impractical to 
empirically study the effect of a multitude of parameters in our optical 
trapping system, we have chosen to model the apparatus using a purpose-built 
three-dimensional Monte Carlo (MC) simulation.

Existing simulations included some, but not all, of the features necessary to 
fully model our apparatus. Early simulations (ca. 1990) typically only 
simulated individual subsystems of an optical trapping apparatus, as the 
available computational resources of the era limited the complexity that could 
be reasonably modeled.  One such MC simulation modeled atomic beam cooling, 
but approximations, such as assuming isotropic spontaneous emission in the 
presence of circularly polarized laser beams, only allowed for upper limits on 
the atomic beam velocity distribution to be obtained~\cite{Blatt1986}.  A 
Langevin-equation based simulation of atoms in one- and three-dimensional 
optical molasses predicted temperatures of cooled Na and Cs atoms in agreement 
with experiments, but also produced exotic results such as non-Gaussian 
velocity distributions~\cite{Javanainen1992}.  As computational power 
increased over the following decades, simulations of the laser cooling and 
trapping process increased in complexity.  A two-dimensional simulation~\cite
{Vredenbregt2003} used multiple cooling and trapping subsystems chained 
together to model experiments, but did not include any additional experimental 
factors, such as the atomic beam source.  A recent simulation~\cite
{Hamamda2015} of the slowing of a supersonic atomic beam incorporated a 
three-dimensional treatment of atoms in a laser-field, but did not include 
atomic beam collimation nor optical trapping.  Fully quantum treatments of 
light-atom interactions~\cite{Castin1995,Dunn2006} used MC wave-function 
techniques, but as with earlier simulations, typically only looked at 
individual subsystems, such as optical molasses.

In the absence of an existing simulation that could fully model our 
\Ra~optical trap, we developed a laser cooling and trapping MC simulation that 
includes both the main subsystems of an optical trapping system and additional 
experimental details of our apparatus such as the atomic beam source.  In 
addition to applying it to our \Ra~optical trap, we benchmarked our simulation 
against a well-characterized \Sr~optical trap used in an optical lattice clock 
at JILA.  The simulation replicates certain performance characteristics of 
both optical traps, such as the relative gain in trapping efficiency from 
transversely cooling the atomic beam, but does not reproduce the measured 
overall trapping efficiency by a factor of 30 for the \Ra-optical trap.

In this work we present the decade-long development of our MC simulation and 
its application to our \Ra~optical trap.  In Sec.~\ref{sec:systems} we 
describe the experimental setups of the \Ra~and \Sr~ optical traps considered 
in the simulation.  Sec.~\ref{sec:simulation} details the features of the MC 
simulation.  We present the simulation results in Sec.~\ref{sec:results} and 
further systematic studies examining the discrepancy between the measured and 
simulated results in Sec.~\ref{sec:systematics}.  Sec.~\ref{sec:conclusion} 
outlines our planned upgrades to and future use of the simulation.  Our 
treatment of light-atom interactions, atomic beam source angular \& velocity 
distributions, and estimates of the relevant van~der~Waals $C_6$ coefficients 
are detailed in Appendices \ref{app:rates}, \ref{app:distros}, and \ref
{app:c6coeff} respectively.

\section{The Optical Trapping Systems}\label{sec:systems}

\subsection{Ra optical trap}\label{sec:ratrap}
The Ra~optical trap was developed and optimized at Argonne National Laboratory 
for use in a next-generation search for a permanent EDM~\cite{Griffith2009,
Holt2010}.  The optical trap is fully detailed in Refs.~\cite{Guest2007,
Parker2012,Parker2015} and we present here an overview of the subsystems 
relevant to our simulation. $^{226}\mathrm{Ra}$ ($\tau_{1/2} = 1600\ \mathrm
{yr}$, nuclear spin~$= 0$) is used for testing and optimizing the trap, while 
$^{225}\mathrm{Ra}$ ($\tau_{1/2} = 15\ \mathrm{d}$, nuclear spin~$= 1/2$) is 
used for the EDM measurements.

\begin{figure}[t]
\includegraphics[width=\columnwidth]{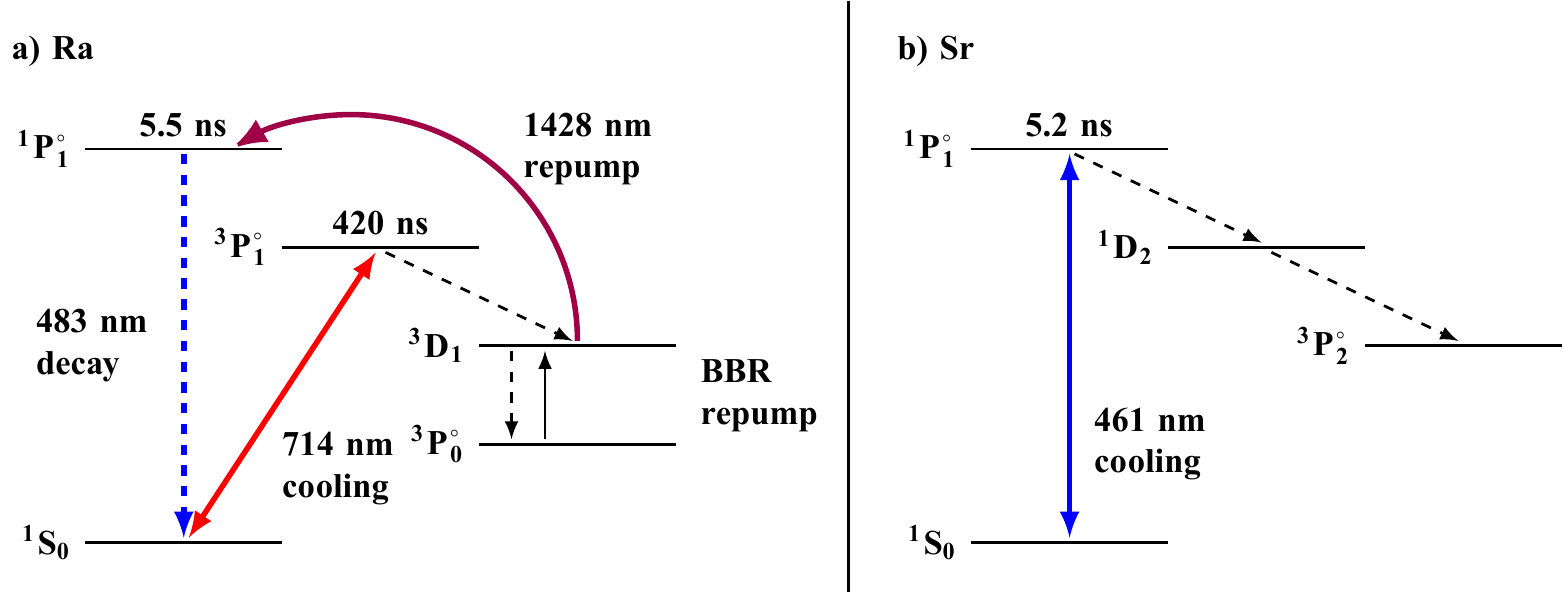}
\caption{Simplified level diagram for Ra (left) and Sr (right). 
For Ra, we handle decays to the metastable states $^3\!D_1$ and $^3\!P_0$  by 
a repump laser and room-temperature blackbody radiation (BBR) as described in 
Ref.~\cite{Guest2007}.  With Sr, decay to metastable states occurs with a time 
constant of 20~ms, although two repump lasers (not shown) permit continued 
cycling on the 461~nm transition. 
 \label{levels}}
\end{figure}

The lowest-lying $^1\!S_0 \leftrightarrow\, ^1\!P_1$ transition is typically 
used for laser cooling of alkaline-earth atoms, e.g.\ Ca and Yb, but we do not 
use this transition for Ra~due to the high probability for an atom to decay 
into the metastable $D$ and $P$ states.  Instead, we use the weaker $7s^2 \, 
^1\!S_0 \leftrightarrow\, 7s7p \,  ^3\!P_1$ transition near $\SI{714}
{\nano\meter}$ for laser cooling and trapping, see Fig.~\ref{levels} (left).  
This intercombination transition is weakly allowed in divalent atoms due to 
singlet-triplet mixing \cite{Breit1933}; for Ra, the $^3\!P_1$ lifetime is 
422~ns \cite{Scielzo2006}, yielding a natural linewidth $\gamma_{\rm{Ra}}/2\pi 
\approx$~380~kHz.  While this transition is sufficiently strong for trapping 
Ra~atoms~\cite{Guest2007}, we are only able to slow atoms with initial 
velocities of up to $\SI{63}{\meter\per\second}$.

\begin{figure}[ht]
\includegraphics[width=\columnwidth]{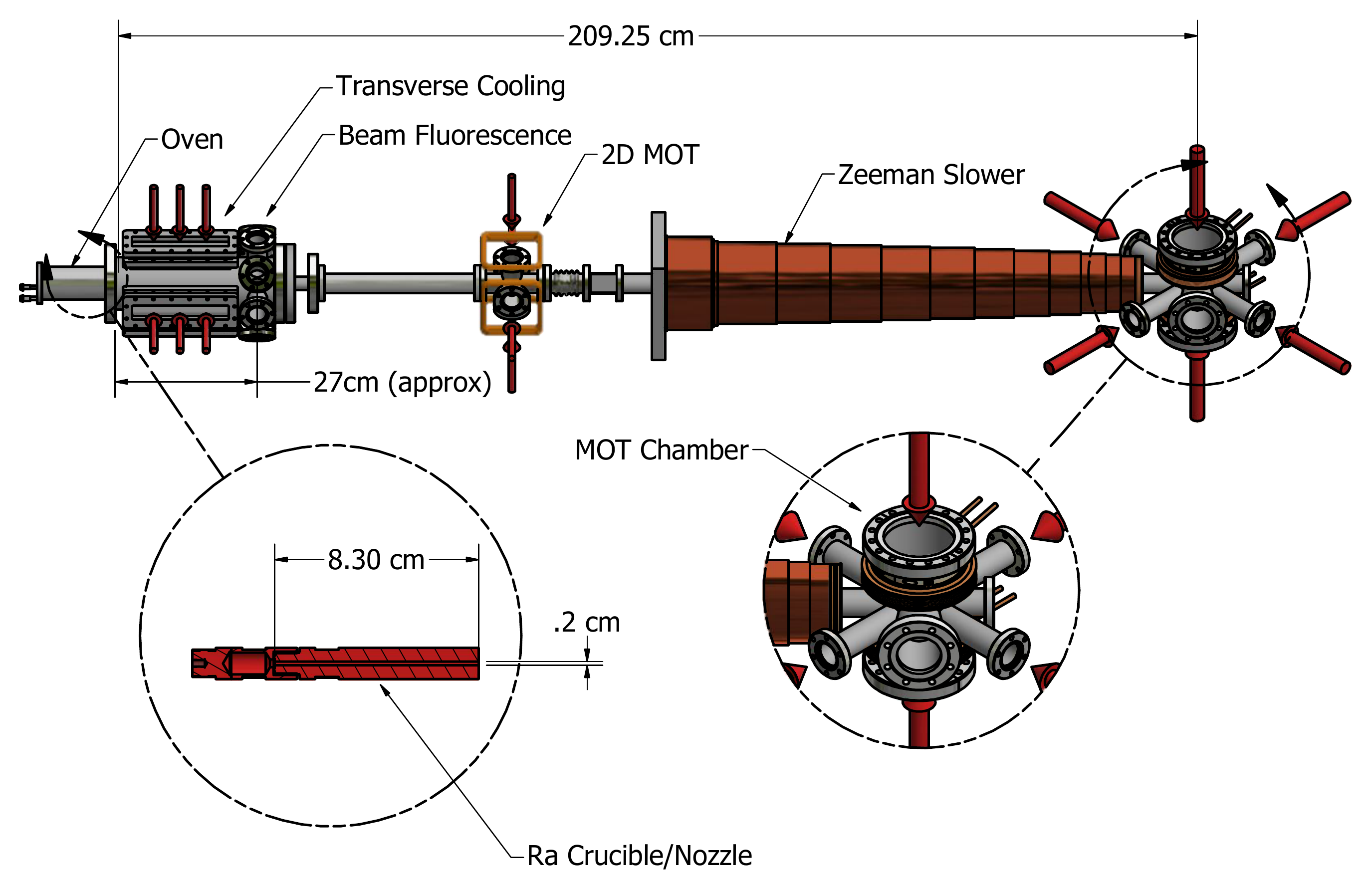}
    \caption{Schematic diagram of the Ra optical trapping system that has been 
    developed at the Argonne National Laboratory.  The inset figures show 
    details of the oven system and the 3D magneto-optical trap (MOT).   The 
    figure is to scale and the overall length from the output of the oven to 
    the MOT is about 2~m.\label{ratrap}}
\end{figure}

A diagram of the laser trapping portion of the apparatus is shown in 
Fig.~\ref{ratrap}.  We heat a crucible containing metallic Ba and Ra(NO$_3$) 
to about $\SI{500}{\celsius}$, sufficient to produce a beam of atomic Ra~
(chemically reduced by the Ba) that we collimate with a nozzle of length $L_
{\text{noz}}=\SI{8.3}{\cm}$ and diameter $d_{\text{noz}}=\SI{0.2}{\cm}$.  At 
this temperature, we determine the total integrated beam flux to be $\SI{5e9}
{\per\second}$ by collecting laser-induced fluorescence on a photomultiplier 
tube from the atomic beam approximately $\SI{27}{\cm}$ downstream from the 
oven's nozzle.  An approximate beam divergence half-angle of $\SI{13}
{\milli\radian}$ is expected based on the aspect ratio of the nozzle in the 
molecular flow limit.

After exiting the nozzle, the atoms are transversely cooled with a 2D optical 
molasses formed by two orthogonal laser beams each making about 10 passes 
across the atomic beam.The atomic beam then passes through a 
conductance-limiting tube with diameter $\SI{2.5}{\cm}$ and length $\SI{1}{cm}
$, see Fig.~\ref{ratrap}.  Subsequently, the atoms enter a 1~m long constant 
deceleration Zeeman slower.  Here, we use a $\sigma^+$ slower configuration 
where the magnetic field decreases with distance as the atoms are slowed.  
This type of slower is appropriate for the weak transition available in 
Ra~because it allows the slowing region to extend all the way to the 
magneto-optical trap (MOT).  Moreover, the atoms are less affected by 
transverse heating during the slowing process.  The Zeeman slower uses a peak 
field of 12~G to capture atoms with velocities up to $\SI{63}
{\meter\per\second}$.  Finally, we use a three-dimensional MOT to trap the 
atoms, where a single repump laser beam near 1428~nm and room-temperature 
blackbody radiation  keep the atoms cycling on the cooling transition, see 
Fig.~\ref{levels} (left) and Ref.~\cite{Guest2007}.

\subsection{\Sr~optical trap}\label{sec:srtrap}
\begin{figure}[ht]
\includegraphics[width=\columnwidth]{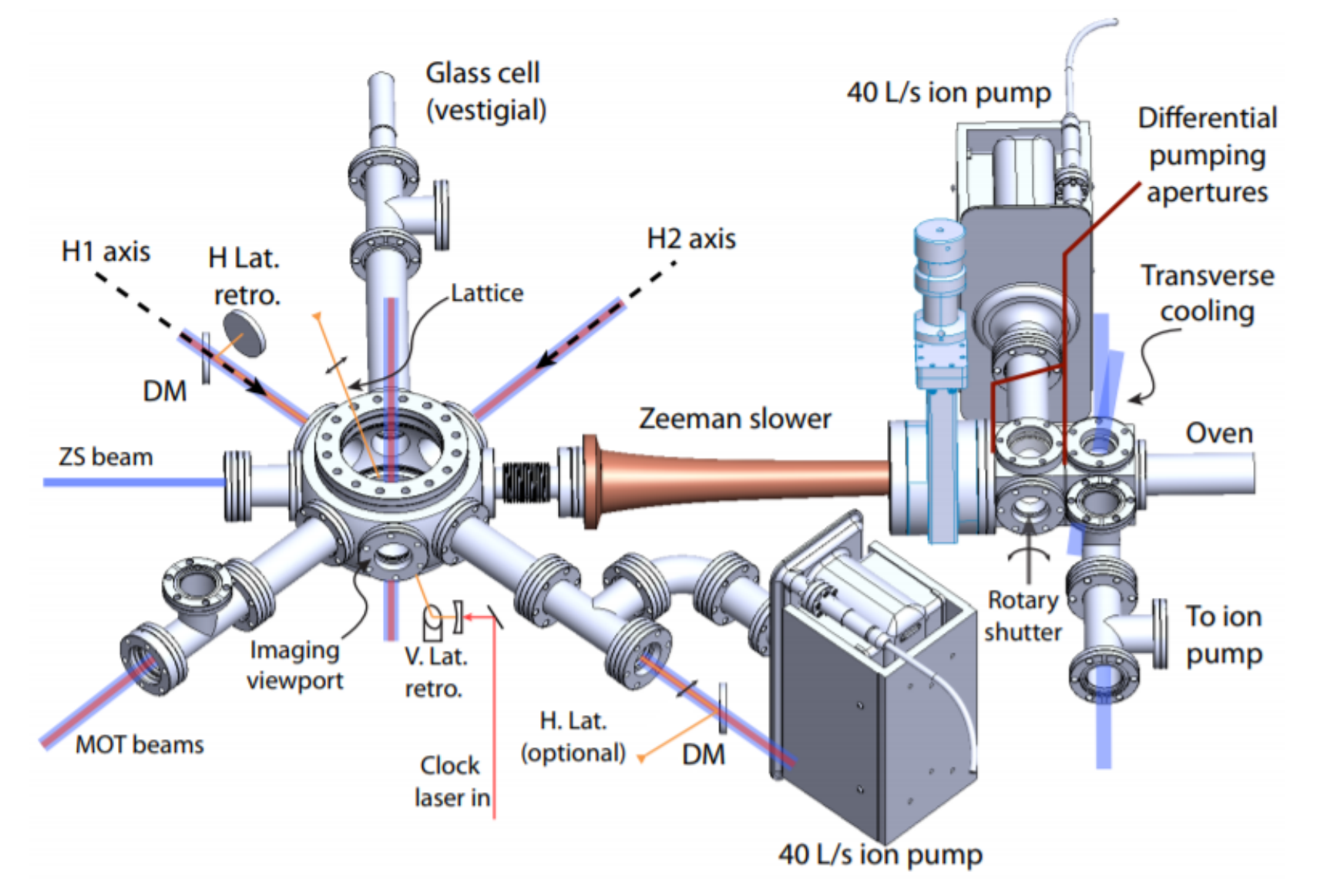}
\caption{\label{fig:srtrap}
The \Sr~optical trapping apparatus.  Figure from Ref.~\cite{Martin2013}.}
\end{figure}

The \Sr~(stable, $\mathrm{nat.\ abd.=0.83}$, nuclear spin~$=0$) system of the 
optical lattice clock developed at JILA~\cite{Ludlow2008} is fully described 
in Refs.~\cite{Loftus2004,Ludlow2008a,Boyd2007,Martin2013}.  Here we focus only on the 
\Sr~source and subsequent atomic beam, transverse cooler, Zeeman slower, and 
first-stage MOT, see Fig.~\ref{fig:srtrap}.  The atomic beam source is an 
effusive oven with a cylindrical nozzle of $\SI{0.2}{\cm}$ diameter and $\SI{2}
{\cm}$ length.  The crucible and nozzle temperatures are $\SI{525}{\celsius}$ 
and $\SI{725}{\celsius}$, respectively. A $\SI{0.36}{\cm}$ diameter filtering 
aperture located $\SI{19}{\cm}$ downstream of the nozzle provides additional 
collimation.

All laser-based interactions utilize the electric dipole transition $5s^2 \,
^1\!S_0 \leftrightarrow\, 5s5p \, ^1\!P_1$ near 461~nm, see Fig.~\ref{levels} 
(right), with a natural decay rate $\gamma_{\rm{Sr}}/2\pi \approx \SI{32}
{\mega\hertz}$~\cite{Loftus2004}.  The first laser-atom interaction region is 
a transverse cooling region located shortly after the filtering aperture; this 
two-dimensional optical molasses is formed by two orthogonal laser beams, each 
retro-reflected and elliptically shaped to have a long axis in the direction 
of atomic beam propagation, maximizing the interaction region.  Next, the 
atomic beam passes through a gate valve and a shutter before entering the 
Zeeman slower.  In this case, the slower is operated in a $\sigma^-$ 
configuration~\cite{Barrett1991}, wherein the magnetic field increases with 
distance as the atom slows.  This facilitates a sharp cutoff in the laser-atom 
interaction as the atom exits the slower.  Here, the slower uses a peak 
magnetic field near 600~G to capture atoms with longitudinal velocities up to 
about $\SI{500}{\meter\per\second}$.  The Zeeman slower laser beam is assumed 
to be collimated with an intensity radius ($1/\rm{e}^2$) of $\SI{0.2}{\cm}$ 
and a saturation parameter of 12 near the center of the Gaussian beam\footnote
{Regarding the beam parameters used in the \Sr~Zeeman slower, we  note \cite
{Ludlow2013} that the beam shape is a distorted Gaussian, which when 
spatially-filtered gave nearly the same slowing efficiency despite having 
approximately half the total power. Thus we take the effective power as $\SI
{30}{\milli\watt}$.}~\cite{Ludlow2013}.

\begin{table}[ht]
\bgroup
\def\arraystretch{1.5}
\begin{tabular}{l c c c}
\toprule Parameter & Units &  Sr expt. & Ra expt. \\
\hline
Cooling transition & & $ ^1\!S_0 \leftrightarrow\, ^1\!P_1$  & $ ^1\!S_0 
    \leftrightarrow\,^3\!P_1$ \\
Transition linewidth $\gamma$ &  MHz & 32 & 0.38  \\
Saturation intensity $I_s$ &  mW/cm$^2$ & 41 & 0.14 \\
Transverse cooling intensity & $I_s$  & 0.3 & 120 \\
Zeeman slower peak field & G &  600 & 12 \\
Zeeman slower  detuning & $\gamma$ &  34  & 10  \\ 
Zeeman slower intensity & $I_s$ & 12  & 80 \\
MOT laser detuning & $\gamma$ & 1.3 & 6 \\
MOT beam intensity  & $I_s$  &  0.04 & 11 \\
MOT B-field gradient & G/cm & 50 & 1 \\
\botrule
\end{tabular}
\egroup
\caption{\label{parameterTable}
Comparison of key parameters in the Sr~and Ra~experiments. All intensities are 
given as the peak intensity of the Gaussian beam, before retroreflection, and 
in units of the saturation intensity. All detunings are given in units of the 
transition linewidth. }
\end{table}

After exiting the Zeeman slower, the slowed atoms are captured by a 
three-dimensional MOT.  In the absence of repumping, the cold atoms eventually 
decay into the metastable ``dark'' state $^3\!P_2$, with a $\SI{20}
{\milli\second}$ time constant.  According to Ref.~\cite{Loftus2004}, the 
loading rate of the MOT is $\SI{2.7(9)e9}{\flux}$ while the atomic flux after 
the filtering aperture is $\SI{3e11}{\flux}$, which gives the efficiency of 
the system (excluding the losses due to beam collimation from the filtering 
aperture) as $\eta_{\text{Sr,exp.}}=\num{9e-3}$.  With two additional lasers 
to repump the atoms shelved in the long-lived $^3\!P_2$ state, the MOT 
lifetime is increased by a factor of 15 and becomes limited by collisions with 
the residual gas and/or the \Sr~atomic beam, which passes directly through the 
MOT.  See Tab.~\ref{parameterTable} for a summary of the key parameters for 
both experiments.

\section{Monte Carlo Simulation}\label{sec:simulation}
In order to fully model the optical traps described in Sec.~\ref{sec:systems}, 
our MC simulation includes the main optical trapping subsystems (transverse 
cooler, Zeeman slower, and MOT) and as well as additional experimental 
details: the atomic beam source, gravity, laboratory magnetic fields, laser 
frequency noise, and residual gas collisions.  We simulate the random walk of 
individual atoms in gravitational free-fall along the length of the modeled 
apparatus.  The simulation outputs both the complete state of sampled atoms at 
each time step as well as the collective statistics of an entire simulation 
run (e.g.\ the overall trapping efficiency).  The following subsections detail 
our implementation of these subsystems and processes.

\subsection{Modelling the Atomic Beam Source}\label{sec:oven}
An effusive oven consists of a crucible and a collimating nozzle.  These ovens 
operate in one of three flow regimes: molecular, intermediate, and viscous.  
These flow regimes determine the angular intensity and velocity distributions 
of the output atomic beam.  The temperature of the crucible and geometry of 
the nozzle determine the flow regime of the oven, which is characterized by 
the Knudsen number,
\begin{align}\label{eq:knudsen}
    K_{n_S} &= \frac{\lambda_{\text{atom}}}{d_{\text{noz}}} & \text{(short wide nozzle)},\nonumber\\
    K_{n_L} &= \frac{\lambda_{\text{atom}}}{L_{\text{noz}}} & \text{(long narrow nozzle)},
\end{align}
where $\lambda_{\text{atom}}$ is the mean free path of an atom in the oven, 
see Eq.~\ref{eq:mfp} in appendix~\ref{app:distros}, and $d_{\text{noz}}$ and 
$L_{\text{noz}}$ are the diameter and length of the nozzle, respectively.  The 
aspect ratio of the nozzle,
\begin{equation}\label{eq:aspect}
    \chi_\mathrm{AR} = \frac{d_{\text{noz}}}{L_{\text{noz}}},
\end{equation}
characterizes the shape of the nozzle: $\chi_\mathrm{AR}>1$ for short wide 
nozzles and $\chi_\mathrm{AR}<1$ for long narrow nozzles.  Both the Ra~and 
Sr~oven have long narrow nozzles, $\chi_\mathrm{AR}\qty(\text{Ra})=\num{2.4e-2}
$ and $\chi_\mathrm{AR}\qty(\text{Sr})=\num{0.1}$, and going forward we will 
refer to the Knudsen number as $K_n \equiv K_{n_L}$.  For molecular flow ($K_n 
\gg 1$), the geometry of the nozzle determines the shape of the angular 
intensity distribution.  In the intermediate flow ($K_n \approx 1$) regime, 
one must account for the influence that both interatomic collisions and the 
geometry of the nozzle have on the shape of the angular intensity 
distribution.  For viscous flow ($K_n \ll 1$), hydrodynamic effects dominate 
the shaping of the angular intensity distribution and the total flow rate from 
the oven.  In molecular and intermediate flows, the velocity distribution of 
effusing atoms is a temperature-dependent Boltzmann distribution, perturbed by 
a Knudsen-number dependent factor for atoms with trajectories originating from 
the crucible (as opposed to a nozzle wall).  See App.~\ref{app:distros} for 
the formulation of the molecular and intermediate flow regime distributions; 
we do not consider viscous flow in the simulation.

We generate the angular intensity and velocity distributions from the geometry,
 temperature, and Knudsen number of the effusive oven in the optical trapping 
system we are modeling (e.g.\ the two optical traps described in Sec.~\ref
{sec:systems}).  A single atom is randomly placed within the area of the exit 
of the nozzle, and its trajectory is randomly sampled from the appropriate 
angular intensity distribution.  We then project this trajectory backwards 
into the nozzle and crucible to determine the last wall with which the atom 
last came in contact. We assume the atom reaches thermal equilibrium before 
leaving this wall, so that the temperature of this wall sets the Boltzmann 
distribution from which we randomly sample the atom's velocity.  Considering 
the crucible and nozzle temperatures separately allows simulation of 
temperature gradients in the oven.

\subsection{Modeling Laser-Atom Interactions}\label{sec:interactions}
The simulation models each laser field as a composition of circularly and 
linearly polarized laser beams.  Each of these beams is defined by a reference 
point, propagation direction, transverse elliptical size, Gaussian intensity 
profile, divergence angle, polarization helicity, saturation intensity, and 
frequency detuning.  We define two types of laser beams in the simulation: 
transition and repump beams.  Each of these beams compete for the excitation 
of simulated atoms.  

The transition beams operate on a defined atomic transition, which we treat as 
a two-level system with hyperfine splitting.  We treat laser-atoms 
interactions semiclassically in the simulation: photon absorption and 
spontaneous emission follow electric dipole selection rules and angular 
distributions, and generate velocity kicks to the atom.  Atoms reside in 
quantum states described by the total electronic angular momentum plus nuclear 
spin quantum number $F$ and its projection $m_F$ along the quantized axis, 
which we set along the direction of the external B-field.  We assume that both 
$F$ and $m_F$ are conserved during ballistic flight and that stimulated 
emission is negligible.  The total photon absorption rate is calculated by 
first determining the Zeeman shift of the $m_F$ states caused by the local 
magnetic field, see Sec.~\ref{sec:bfields} below, and then summing over the 
Doppler-shift dependent saturation intensity of each beam.  We take the photon 
absorption rate of any one laser beam as the total absorption rate weighted by 
the fractional saturation intensity contributed by that beam.  See App.~\ref
{app:rates} for further details on the formulation of these rates.

We also include a stochastic noise process for transition beams such that the 
central frequency of a beam can undergo a random walk before being damped back 
to the central value.  The noise is specified by an amplitude $\sigma$ and 
damping time $\tau$.  The physical significance of the damping time can be, 
for example, a laser frequency servo loop with finite bandwidth, which in the 
Ra~optical trap gives a damping time of $\tau=\SI{50}{\micro\second}$.  The 
corresponding spectral density at each Fourier frequency $f$ is white from 
D.C. to $f = 0.5/\tau$, after which it falls off like $1/f$.

We treat the repump beams and dark states more simply.  Atoms in the excited 
state can decay to a metastable dark state.  We model these dark states with a 
transition probability from the excited state and a characteristic lifetime 
that an atom will remain in the dark state.  Multiple dark states are treated 
cumulatively and modeled as a single dark state with an overall transition 
probability and characteristic lifetime.  If an atom is in a dark state and 
within the defined bounds of a repump laser beam, we return the atom to the 
ground state after a randomly-selected lifetime-weighted time step.

\subsection{Magnetic Fields}\label{sec:bfields}
Magnetic fields play a key role in modeling subsystems of an optical trapping 
apparatus, specifically the Zeeman slower and MOT.  We include magnetic fields 
in the simulation from multiple sources: circular coil windings with specified 
current, ambient field maps, constant background fields, and a two-dimensional 
axisymmetric Zeeman slower.  From these we calculate the total magnetic field 
as a three-dimensional grid of magnetic field vectors.  The value of the local 
magnetic field is found by interpolating between points on this grid.

\subsection{Collisions With Residual Gas Molecules}\label{sec:bggas}
We consider the effect of collisions with residual gas molecules to simulate 
how a non-zero vacuum pressure alters an atom's trajectory in the simulation.  
Due to shallow trap depths, long-range dispersion forces in the form of 
induced dipole-dipole force $V = -C_6/R^6$ dominate the collisional cross 
section for ejection from a MOT, where $C_6$ is the long-range van der Waals 
coefficient and $R$ is the intermolecular distance.  To determine the 
collision frequency, scattering angle, and outgoing velocity, we first 
calculate the mean free path of the atom based on kinetic theory, from which 
we randomly determine the distance the atom travels before a collision.  When 
a collision event occurs, we randomly sample the velocity of the residual gas 
molecule from a room-temperature Boltzmann distribution, and an impact 
parameter from a uniform distribution.  Collisions with impact parameters 
beyond a specified cutoff value are ignored as we assume these lack the energy 
to significantly alter an atom's trajectory.  With these values, and specified 
$C_6$ coefficients and interaction distances, we numerically calculate the 
Lennard-Jones dynamics~\cite{Venka2012} to find the scattering angle and 
outgoing velocity, and update the atom's trajectory accordingly.

\subsection{Simulating An Atom's Random Walk}\label{sec:ranwalk}
\begin{figure}[ht]
\includegraphics[width=\columnwidth]{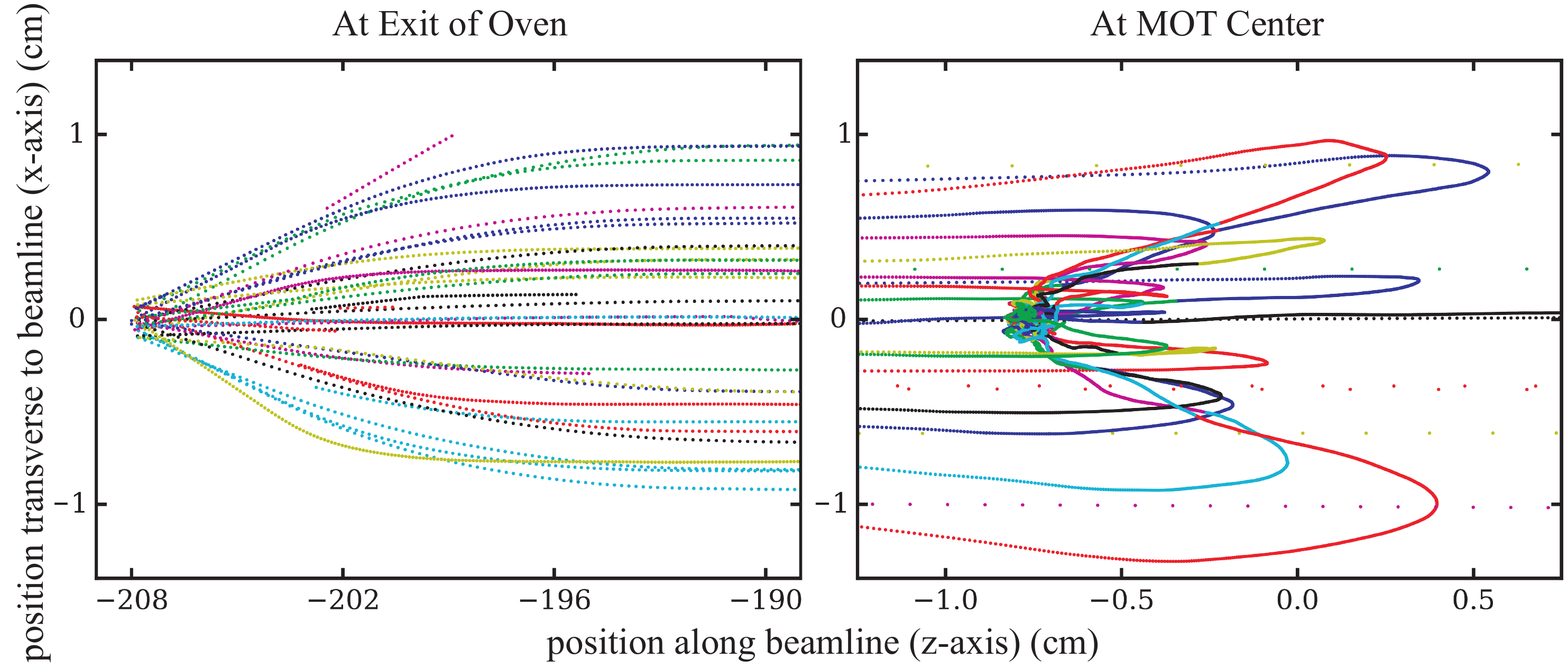}
    \caption{Examples of atomic trajectories for the Radium Optical Trap. Note 
    that the vertical scale is the same for both, but the 
    horizontal scales are different.\label{fig:atomtraj}}
\end{figure}

We begin the simulation by generating an atom at the source oven as described 
in Sec.~\ref{sec:oven}, see Fig.~\ref{fig:atomtraj} (left).  The atom will 
then proceed in a random walk through a series of time steps chosen based on 
the photon scattering and residual gas collision rates.  All time steps are 
clamped to a specified maximum value.  At the start of a time step, we 
calculate the photon absorption, spontaneous emission, repumping, and residual 
gas collision rates.  Based on these rates, the shortest time step until the 
next event is chosen.  We then update the atom's position and velocity 
accordingly while holding these rates constant.  At the end of the time step, 
the process indicated by the rate the time step is chosen from is performed, 
and then a new iteration is started.  We continue tracking an atom through 
these time steps until one of the following scenarios occur: the atom 
encounters a physical barrier (such as an aperture), leaves the defined bounds 
of the simulation, exceeds the specified total time limit, or becomes trapped 
in the MOT, see Fig.~\ref{fig:atomtraj} (right).  Finally, we terminate the 
simulation of that atom and update the lost and trapped atom counts 
accordingly.
\subsection{Numerical Implementation}
The simulation code consists of modules written in Fortran 90 and C, and we 
utilize the CERNLIB Programming Library~\cite{cernlib} for numerical 
calculations.  In order to parallelize our code, we use the Master/Slave 
algorithm implemented by the Asynchronous Dynamic Load Balancing (ADLB) 
Fortran interface~\cite{Bogner2013,Lusk2010} to Message Passing Interface 
standard~\cite{MPI}.  Using ADLB allows for work-sharing among a number of 
processors and computer nodes in a distributed-memory system.  This permits 
rapid investigation of a large number of parameters that affect the efficiency 
of the simulated optical traps.

\section{Comparison to Experiment}\label{sec:results}
In order to accurately model each of the optical traps described in Sec.~\ref
{sec:systems}, we include all laser beams, known magnetic fields, and geometry 
using experimentally determined parameters.  For each simulation, initial 
velocity and angular cuts discard atoms with velocity vectors that immediately 
leave the simulation bounds upon exiting the oven and atoms that have an 
initial velocity exceeding what the Zeeman slower can sufficiently slow.  We 
minimize statistical errors in the simulation by simulating several thousand 
atoms within these angular and velocity cuts.  In the Ra~experiment we define 
the total efficiency as the ratio of the number of atoms trapped to the number 
of atoms exiting the source oven, as obtained by fluorescence measurements of 
the MOT and atomic beam, see Fig.~\ref{ratrap}.  The \Sr~efficiency definition 
differs in that the number of trapped atoms is compared to the number of atoms 
passing through the filtering aperture, not the total number exiting the oven.

We determine the simulated optical trapping efficiency in the same manner as 
the experimental value.  The fluorescence measurements rely on limited 
knowledge of scattering rates, detector efficiencies, and detector solid 
angles, which results in roughly a factor of 3 uncertainty in the 
experimentally-determined overall trapping efficiencies.  We expect that 
ratios of experimentally-determined trapping efficiencies to be more robust.

For the \Sr~simulation, we use the experimental slowing beam parameters, see 
Sec.~\ref{sec:srtrap} and Tab.~\ref{parameterTable}, and find a \Sr~optical 
trapping efficiency of $\eta_{\text{Sr,sim}}=\num{4e-3}$.  By adding 
transverse cooling, we see a four-fold increase in efficiency to $\eta_{\text
{Sr,sim}}=\num{1.6e-2}$, which is in within a factor of 3 to the experimental 
value of $\eta_{\text{Sr,exp}}=\num{0.9e-2}$.  Additionally, we find the 
trapping efficiency strongly depends on the slowing beam size and intensity.  
We observe that for a fixed optical power, we achieve optimal slowing using a 
central intensity of $4I_s$, and that increasing the beam size while 
maintaining this central intensity leads to higher trapping efficiency due to 
better overlap between the atomic and slowing laser beams.  By expanding the 
slowing beam's intensity radius ($1/e^2$) to $\SI{0.4}{\cm}$, we further 
increase in the \Sr~optical trapping efficiency by another factor of about 
$\num{4}$.

For the \Ra~simulation, we find a baseline optical trapping efficiency from 
slowing of the atomic beam of $\eta_{\text{Ra,sim}}=\num{4e-7}$.  With the 
addition of both transverse cooling and repumping of the atomic beam, we find 
gains in efficiency of approximately 60 and 2.4, respectively, increasing the 
efficiency to $\eta_{\text{Sr,sim}}=\num{5.7e-5}$.  This disagrees with the 
experimental value of $\eta_{\text{Ra,exp}}=\num{2e-6}$ by a factor of $\num
{30}$.  Despite this discrepancy between the experimental and simulated total 
optical trapping efficiencies, we note the relatively good agreement on gain 
in efficiency produced by inclusion of the the transverse cooling region and 
the $\SI{1428}{\nano\meter}$ repump laser (co-propagating with the slower 
laser).  See Tab.~\ref{efficiencyTable} for a comparison of the experimental 
and simulated optical trapping efficiencies in the \Sr\ and \Ra~optical traps.

\begin{table}[ht]
\bgroup
\def\arraystretch{1.5}
\begin{tabular}{l c c c c}
\toprule & \Sr~exp. & \Sr~sim. & \Ra~exp. & \Ra~sim. \\
\hline
$\eta$, ZS only & $\num{2e-3}$ & $\num{4e-3}$ & $\num{1e-8}$ & $\num{4e-7}$ \\
TC gain & $\num{4}$ & $\num{4}$ & $\num{60}$ & $\num{60}$ \\
Repump gain & n/a &n/a & $\num{3.5}$ & $\num{2.4}$ \\
\hline
$\eta$, total  & $\num{9e-3}$ & $\num{1.6e-2}$ & $\num{2e-6}$ & $\num{5.7e-5}$ \\
\botrule
\end{tabular}
\egroup
\caption{\label{efficiencyTable}
Comparison of \Sr~and \Ra~optical trapping efficiencies from experiment and simulation.  ($\eta$: trapping efficiency; ZS: Zeeman slower; TC: transverse cooling) The systematic uncertainty in the experimentally determined total efficiencies is about a factor of 3. 
The statistical uncertainty in the simulated total efficiencies is about 5\%.}
\end{table}

\section{Sensitivity Studies of the Radium Optical Trap}\label{sec:systematics}
To diagnose the source of the discrepancy between the simulated and measured 
values of the \Ra~optical trapping efficiency, we simulate possible systematic 
effects in the experiment that we did not originally account for in the 
simulation.  The simulation's accurate prediction of relative gains in optical 
trapping efficiency allows for sensitivity studies to quantify the extent to 
which these systematic effects contribute to a reduced efficiency.

\subsection{Influence of Residual Gas Inside the Oven}
In both the experiment and simulation, we assume that the the dominant factor 
determining the mean free path of Ra~is the saturated vapor pressure of Ba in 
the oven; all other known materials in the oven have vapor pressures 
significantly lower than Ba at the $\SI{500}{\celsius}$ operating temperature 
($\mathrm{SVP}_\mathrm{Ba} = 138\ \mu\mathrm{Torr}$).  This gives a mean free 
path of $\lambda_{\text{Ra}}\approx\SI{42}{\cm}$ and Knudsen number 
$K_n\approx\num{5.1}$, so we assume the oven operates in the intermediate flow 
regime.  Any additional constituents within the oven with comparable (or 
higher) vapor pressures than Ba will reduce $\lambda_{\text{Ra}}$ and broaden 
the angular distribution of the atomic beam exiting the oven.  We fabricate 
the oven's crucible from Ti, which can absorb and store a significant amount 
of H$_2$~\cite{Yildirim2005}, so we cannot rule out the possibility that there 
is a significant H$_2$ partial pressure.

\begin{figure}[ht]
\includegraphics[width=\columnwidth]{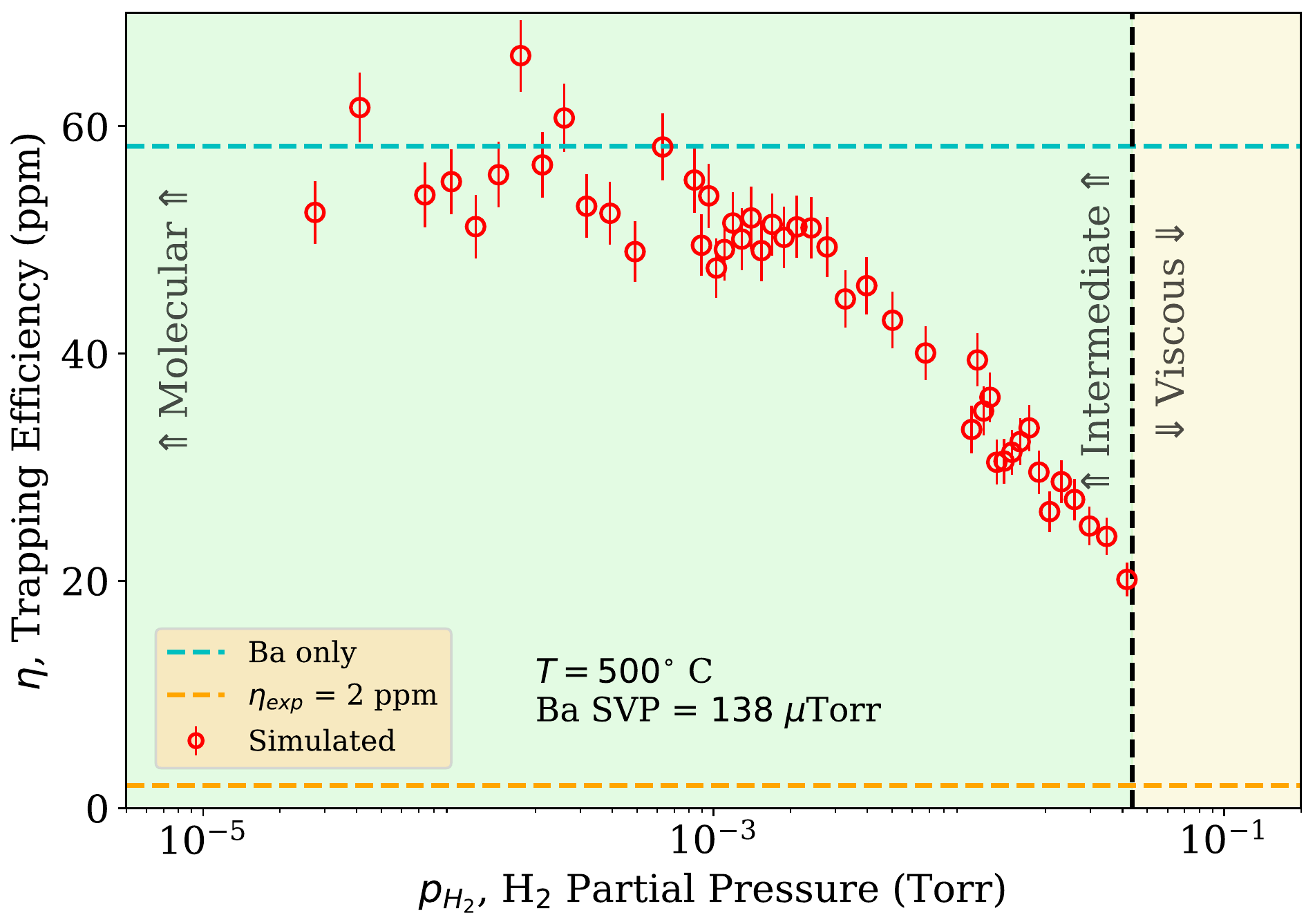}
\caption{\label{fig:knudsen}
The influence of H$_2$ partial pressure in the oven on the optical trapping 
efficiency.  Data points represent a varying H$_2$ partial pressure and a 
constant Ba partial pressure of $\SI{138}{\micro\torr}$ at an oven temperature 
of $\SI{500}{\celsius}$.  The optical trapping efficiency remains unchanged 
until reaching a H$_2$ partial pressure of about $\SI{3e-4}{\torr}$, 
approximately a Knudsen number of $K_n \approx \num{2}$ which is well within 
the bounds of the intermediate flow regime.}
\end{figure}

To characterize the impact that an increased H$_2$ partial pressure in the 
oven has on the optical trapping efficiency, we simulate a varying oven 
angular distribution by varying its Knudsen number from the oven's 
$K_n\approx\num{5.1}$ down to the intermediate/viscous flow regime boundary at 
$K_n = \chi_{\text{AR}} = \num{2.4e-2}$.  We do not consider the viscous flow 
regime in our simulation.  From the results of these simulations, see Fig.~\ref
{fig:knudsen}, we find that it takes a H$_2$ partial pressure in the oven $>\SI
{1e-3}{\torr}$ ($K_n \approx \num{1}$) to see a significant decrease in the 
optical trapping efficiency below our initial results using only Ba to 
determine the oven's angular distribution.  For H$_2$ in the oven to 
significantly contribute to the discrepancy between measured and simulated 
trapping efficiencies, a partial pressure $\gg\SI{2}{\milli\torr}$ is required,
 which may be possible due to differential pumping given the conductance of 
the oven nozzle.  Without means to directly measure the pressure in the oven, 
we cannot rule out the influence of residual gas inside the oven as a 
contribution to the discrepancy between the measured and simulated optical 
trapping efficiencies.

\subsection{Misaligned Oven Nozzle}
\begin{figure}[ht]
\includegraphics[width=\columnwidth]{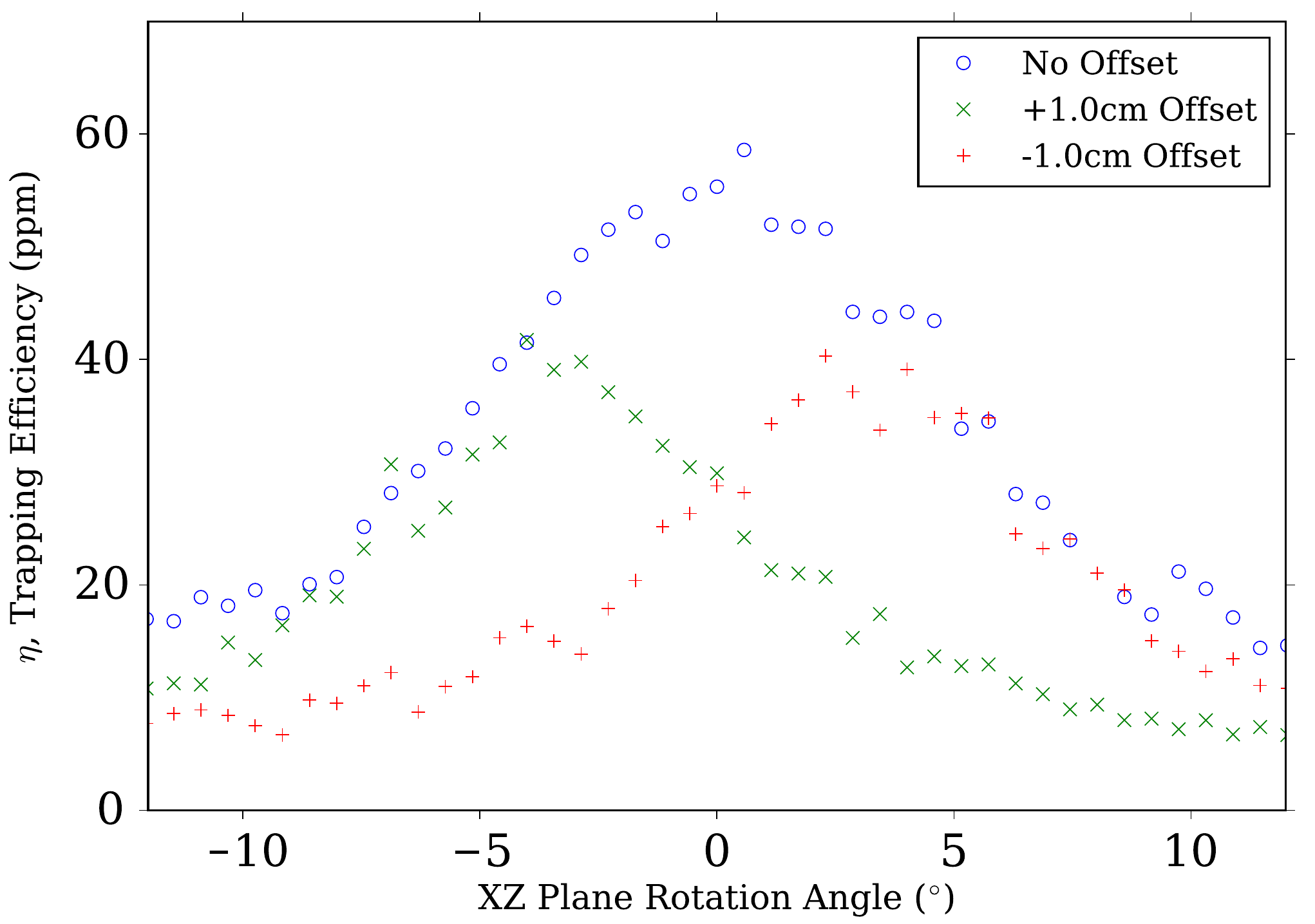}
\caption{\label{fig:nozzle}
Simulated optical trapping efficiency with oven nozzle misalignments.  
Rotations and offsets are in the XZ-plane where the Zeeman slower is aligned 
along the Z-axis.}
\end{figure}

A misalignment between the oven nozzle and Zeeman slower axis potentially 
reduces the optical trapping efficiency.  We can partially correct for this in 
the experiment by carefully realigning the transverse cooling beams.  To model 
these misalignments in the simulation, we rotate and translate the oven nozzle 
off the Zeeman slower axis.  From mechanical constraints, we bound these 
misalignments to $\pm\SI{10}{\degree}$ and $\pm\SI{1}{\cm}$, respectively.  
For these studies we do not modify the properties of the transverse cooler to 
correct for the induced misalignments.  The results, see Fig.~\ref{fig:nozzle},
 show that small misalignments of the nozzle do not alone cause the factor of 
$\num{30}$ discrepancy in trapping efficiency with the experimental results, 
although we cannot rule out a small contribution.

\subsection{Influence of Residual Gas Outside the Oven}
\begin{figure}[ht]
\includegraphics[width=\columnwidth]{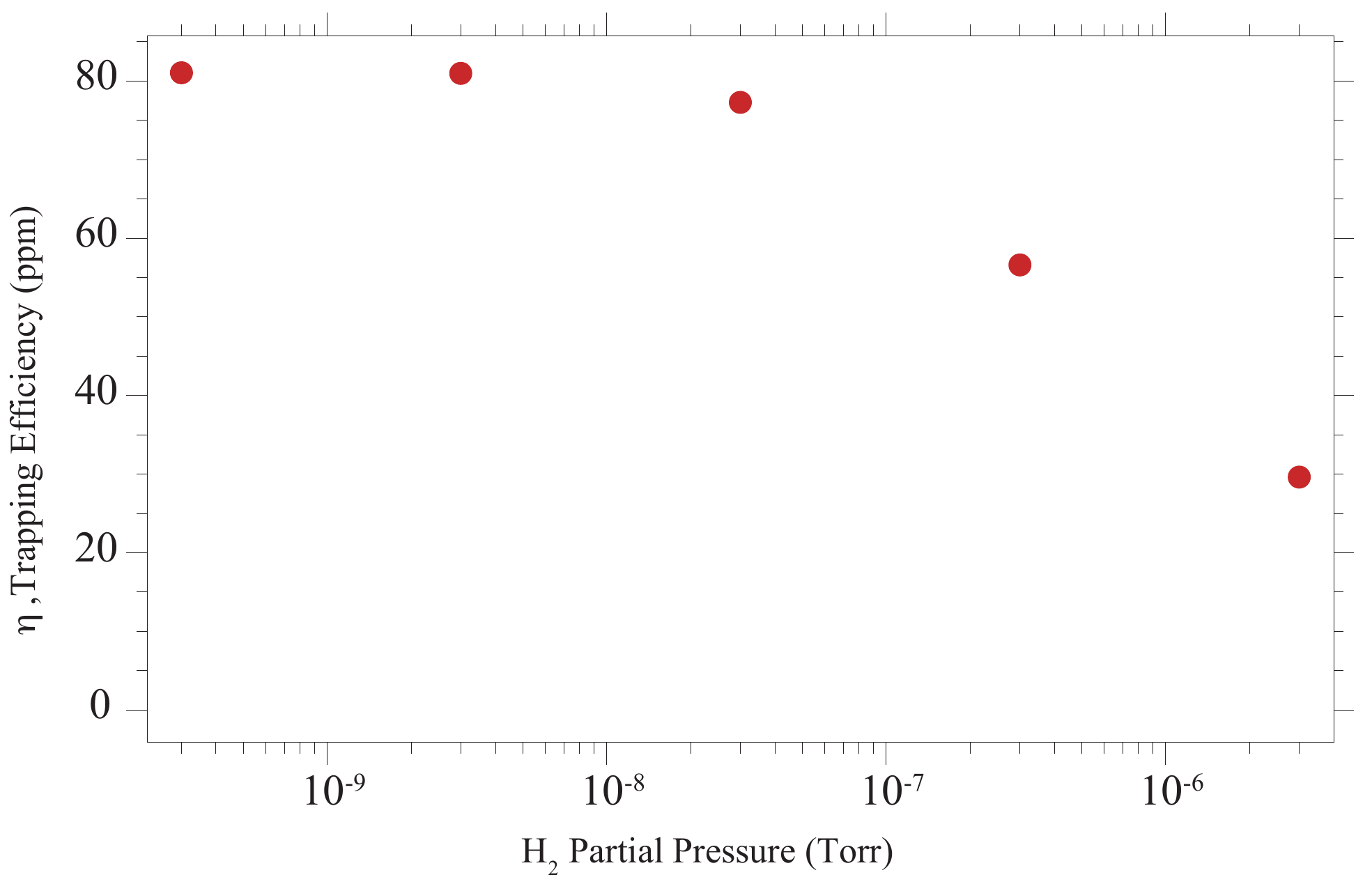}
\caption{\label{fig:rgp}
The influence of vacuum pressure in the transverse cooling region on the 
optical trapping efficiency.  Data points represent a varying $\mathrm{H}_2$ 
number density at room temperature.}
\end{figure}

To test the effect of a non-zero vacuum pressure in the transverse cooling 
region, we simulate collisions between Ra~and a residual gas of H$_2$.  With 
estimates for the $C_6$ coefficients and bond lengths as described in App.~\ref
{app:c6coeff}, we find that our baseline vacuum pressure ($\SI{3e-7}{\torr}$) 
reduces the \Ra~optical trapping efficiency by a factor of 1.4 when compared 
to the collision-free scenario.  Increasing the pressure in the transverse 
cooling region by a factor of 10 further reduces the efficiency by a factor of 
1.9, see Fig.~\ref{fig:rgp}.  These results suggest that the \Ra~optical 
trapping efficiency is only weakly sensitive to residual gas collisions in the 
transverse cooling region, which may indicate that low impact collisions can 
be corrected for with optical forces.  We also simulated the MOT with our 
residual gas collision model and determined a MOT trap lifetime of 15 s which 
agrees with the experimentally determined value at a residual gas pressure of 
$10^{-9}$\ Torr to better than a factor of two.

\subsection{Stray Magnetic Fields in the Transverse Cooler}
\begin{figure}[ht]
\includegraphics[width=\columnwidth]{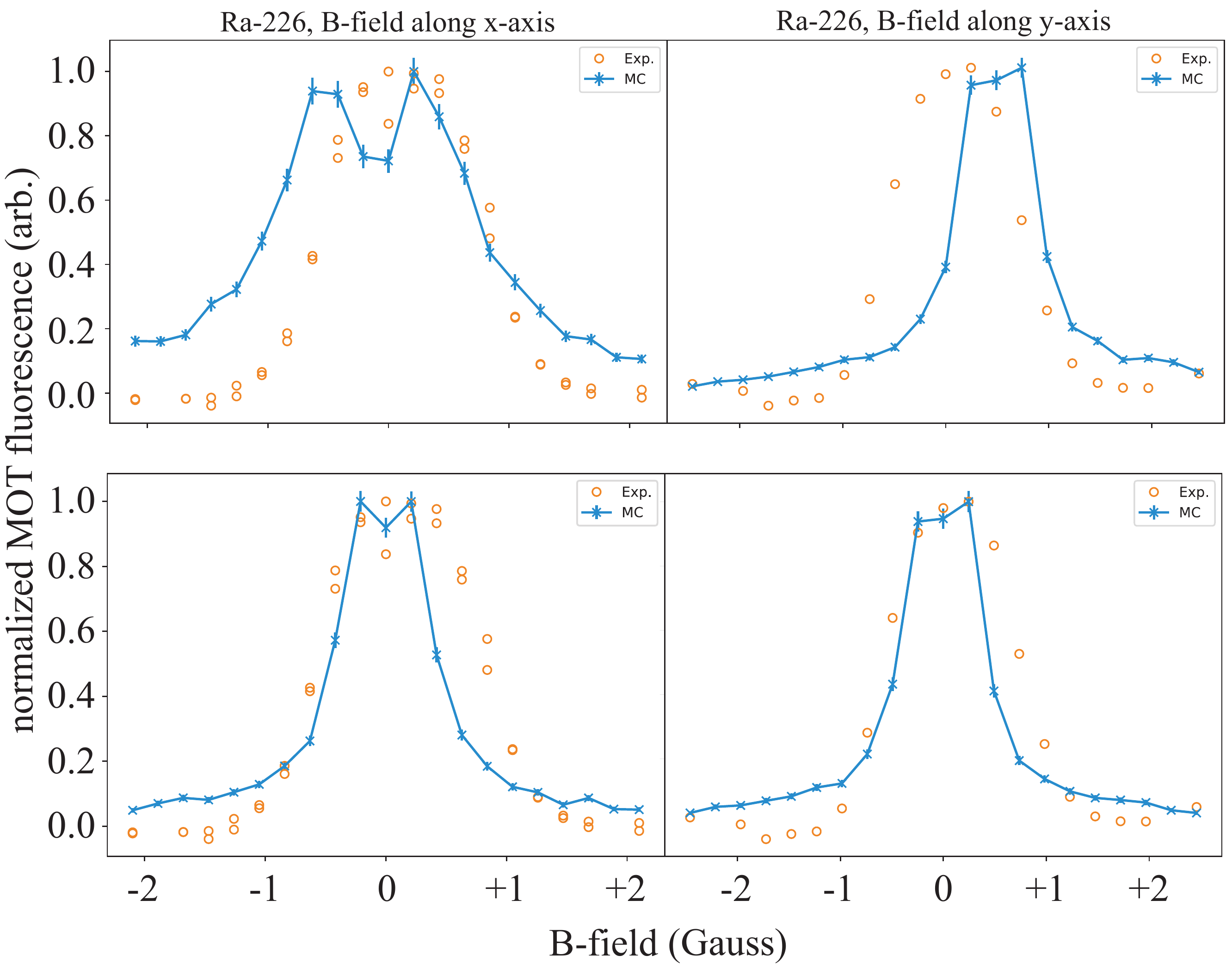}
\caption{\label{fig:bfields}
The influence of stray magnetic fields on the $^{226}$Ra~optical trapping 
efficiency.  Fields are simulated with coil windings placed along each of the 
coordinate axes in the transverse cooling region with varying values of 
current.  The shifts in the peak efficiency are, shown in the upper two plots, 
are due to the Earth's magnetic field in the simulation, which is ``tuned 
out'' in the experimental setup. The bottom two plots show the efficiencies 
when the Earth's magnetic field is ``turned off'' in the simulation.}
\end{figure}

We include all known magnetic sources in the simulation, including the Zeeman 
slower, MOT coils, and the uniform contribution from the earth's magnetic 
field as measured in the laboratory.  We do not explicitly include stray 
magnetic fields, such as those from ion pump magnets and high current heating 
elements in the oven.  In order to test the effect that these stray fields 
have on the trapping efficiency, we model magnetic fields in the transverse 
cooling region by adding fictitious coils and varying their orientation and 
magnetic strength.  The results of these simulations, see Fig.~\ref
{fig:bfields}, show that a stray magnetic field of magnitude in excess of 1 G 
is required to significantly reduce the optical trapping efficiency to the 
levels measured in the apparatus.  Furthermore, in order to test the 
predictions of the simulation results, we added rectangular coils in the TC 
region and experimentally measured the change in the atomic fluorescence from 
the MOT as a function of applied magnetic field in the TC region using $^{226}
\mathrm{Ra}$ as a surrogate for \Ra.  The measured results qualitatively agree 
with the simulation, namely a stray field of in excess of 1 G is needed to 
significantly lower the trapping efficiency.  We do not expect uncontrolled 
fields of this magnitude within the Ra~optical trap so we do not expect these 
stray magnetic fields to be a major contributor to the discrepancy between the 
measured and simulated optical trapping efficiencies.

\subsection{Laser Frequency Noise}
\begin{figure}[ht]
\includegraphics[width=\columnwidth]{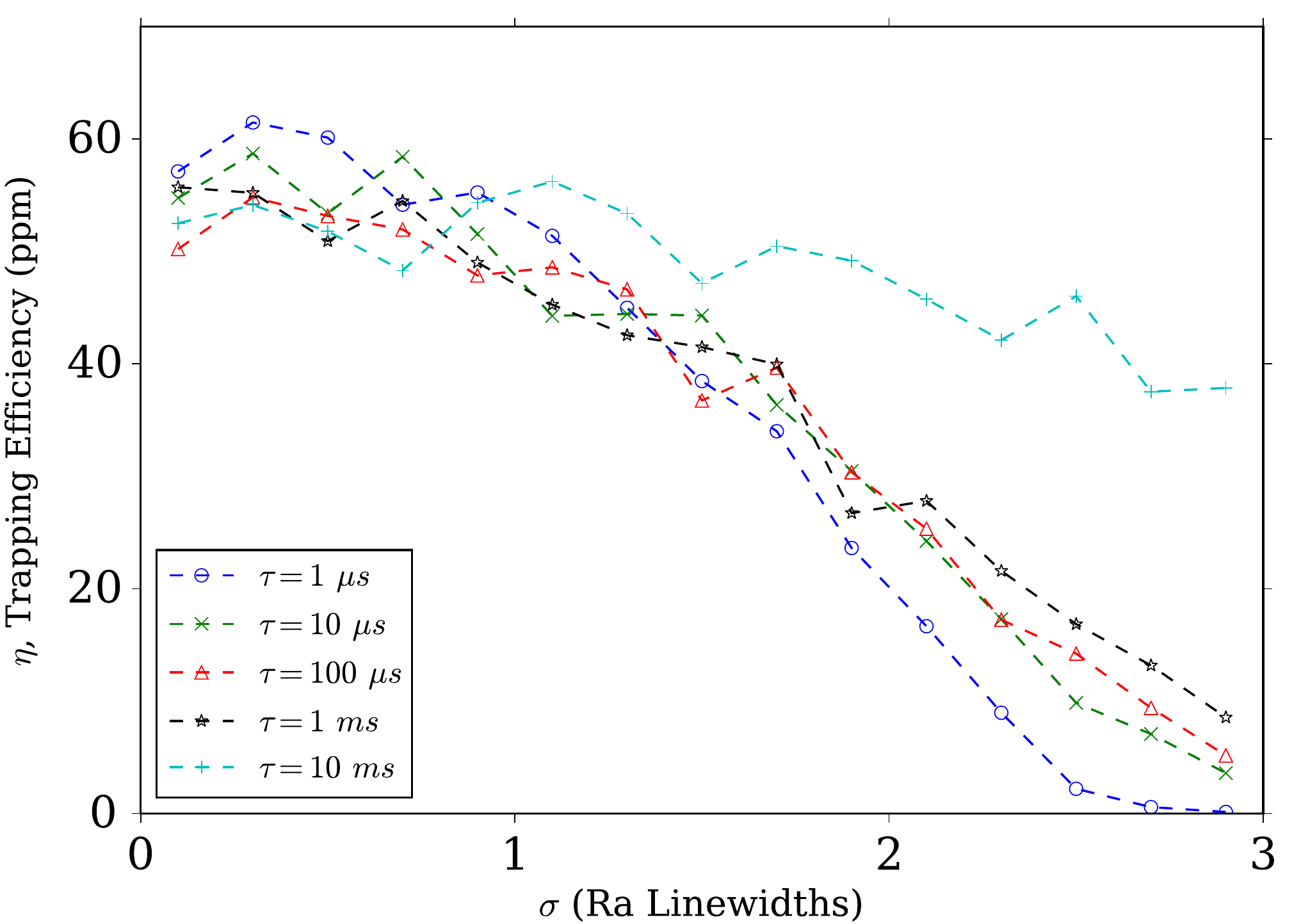}
\caption{\label{fig:noise}
The influence of laser frequency noise on the \Ra~optical trapping efficiency. 
The horizontal axis ($\sigma$) is the laser linewidth in terms of the Ra 
atomic linewidth. The different curves correspond to different values of the 
damping time $\tau$. }
\end{figure}

In the Ra~optical trap, we generate the $\SI{714}{\nano\meter}$ light with a 
Ti:sapphire ring laser referenced to a high-finesse optical cavity.  The 
resulting laser spectrum is mostly white with a few peaks in the $\si
{\kilo\hertz}$ range, and we measure the overall linewidth to be approximately 
$\SI{100}{\kilo\hertz}\approx\SI{0.3}{\ralw}$.  We simulate potential noise 
sources by varying the noise amplitude and damping time associated with 
individual laser beams.  Based on the results of these simulations, see 
Fig.~\ref{fig:noise}, we do not expect the \Ra~optical trapping efficiency to 
be limited by laser linewidth and frequency noise.

\subsection{MOT Beam Misalignment and Imbalance}
We trap Ra~atoms using a 3D MOT created by a quadrupole magnet and three 
orthogonal laser beams of the appropriate helicity intersecting where the 
magnetic field goes to zero.  These beams are retroreflected outside of the 
vacuum chamber.  We measured the optical losses due to reflections from the 
vacuum viewports and found that the retroreflected beams have $\SI{10}
{\percent}$ less power than their forward-going counterparts.  These six laser 
beams are modelled as three intensity-imbalanced orthogonal pairs of 
anti-parallel and opposite helicity beams.  In the simulation, this imbalance 
produces a small shift in the position of the MOT's center, but no change in 
the trapping efficiency is observed.  Further increases to the intensity 
imbalance between pairs of retroreflected beams likewise shifted the MOT 
center while leaving the trapping efficiency unchanged.  This agrees with our 
assumption that the MOT operates at a very high saturation parameter, 
lessening the effect of small changes in the global intensity on the optical 
trapping efficiency.

While ensuring anti-parallelism between each MOT beam and its retroreflection 
is relatively straightforward experimentally, it is more difficult to ensure 
that the global overlap of the three beams is optimal.  We tested the impact 
of sub-optimal MOT beam overlap in the simulation by individually offsetting 
one pair of beams from the others.  Generally speaking, displacements of up to 
$\SI{1}{\cm}$ (approximately the intensity radius of the MOT beams) produces 
no significant losses in the optical trapping efficiency, and we do not expect 
misalignments in the experiment to be any larger than this.

\section{Summary \& Conclusions}\label{sec:conclusion}
We have presented a three-dimensional MC simulation of an optical trapping 
apparatus that includes the effects of several common perturbations to the 
laser cooling and trapping process.  With this simulation we model the optical 
trapping apparatus used in our \Ra~EDM measurements in an effort to improve 
our optical trapping efficiency of \Ra.  Our simulation mostly agrees with the 
experimental results of both the \Ra~and \Sr~optical trapping systems, i.e.~we 
see the same relative gains in trapping efficiency from switching on/off 
components of the apparatus, and we find good agreement between the measured 
and simulated \Sr~optical trapping efficiencies.  However, the \Ra~simulation 
results differ from experimental results by a factor of $30$.  We note that we 
may not see the discrepancy between measured and simulated trapping 
efficiencies in the \Sr~optical trap since we do not consider the full 
trapping efficiency in the system.  Simulations of possible systematic effects 
in the apparatus so far have shown that there is no sole-source of this 
discrepancy, but rather these effects may cumulatively explain this difference 
between the measured and simulated trapping efficiencies.

In addition to these systematic effects, potential contributions to this 
discrepancy may result from other aspects of our simulation.  For example, our 
treatment of the effusive-oven atomic beam source may not be adequate for the 
complex chemistry in the Ra~oven.  We plan on determining the adequacy of this 
atomic beam source model by detailed study of the output of effusive ovens 
with chemical surrogates for Ra (such as Ca) in comparison to the simulation's 
predictions.

While a discrepancy exists between the measured and simulated optical trapping 
efficiencies, the accurate relative gains predicted by the simulation for both 
the Ra~and~\Sr~optical traps makes this simulation a valuable tool for 
studying and optimizing the optical trapping of rare isotopes.  In the case of 
the \Ra~optical trap, we plan to use the simulation to test and optimize and a 
planned ``blue slower'' upgrade.  Our present ``red-slowing'' scheme ($^1\!S_0 
\leftrightarrow\, ^3\!P_1$ near $\SI{714}{\nano\meter}$) can only slow and 
trap atoms with initial velocities of up to $\SI{63}{\meter\per\second}$, 
approximately $<\SI{0.5}{\percent}$ of the oven's velocity distribution.  The 
blue-slower upgrade will use the stronger $^1\!S_0 \leftrightarrow\, ^1\!P_1$ 
transition near $\SI{483}{\nano\meter}$, allowing us to slow and trap atoms 
with initial velocities of up to $\SI{310}{\meter\per\second}$, approximately 
$\SI{50}{\percent}$ of the oven's velocity distribution.  Implementing this 
scheme will require frequency chirping of the slowing laser and additional 
repump lasers to handle the additional atomic dark states associated with this 
transition.  We plan on upgrading the simulation to include both frequency 
chirping and additional repumping transitions, and using the upgraded 
simulation to help implement and optimize the blue slower in the \Ra~optical 
trap.  This upgrade to the \Ra~optical trap will offer increased EDM 
measurement precision by increasing our optical trapping efficiency by up to 
two orders of magnitude.

\begin{acknowledgments}
The authors wish to thank A.~Ludlow for helpful discussions regarding the 
\Sr~apparatus.  This work is supported by the U.S. DOE, Office of Science, 
Office of Nuclear Physics under contracts DE-AC02-06CH11357 and DE-SC0019455, 
the Director's Research Scholars Program at the National Superconducting 
Cyclotron Laboratory, and by Michigan State University through computational 
resources provided by the Institute for Cyber-Enabled Research.  SAF 
acknowledges support from the United States Department of Energy through the 
Computational Science Graduate Fellowship, grant number DE-SC0019323.
\end{acknowledgments}


\begin{thebibliography}{51}%
  \makeatletter
  \providecommand \@ifxundefined [1]{%
   \@ifx{#1\undefined}
  }%
  \providecommand \@ifnum [1]{%
   \ifnum #1\expandafter \@firstoftwo
   \else \expandafter \@secondoftwo
   \fi
  }%
  \providecommand \@ifx [1]{%
   \ifx #1\expandafter \@firstoftwo
   \else \expandafter \@secondoftwo
   \fi
  }%
  \providecommand \natexlab [1]{#1}%
  \providecommand \enquote  [1]{``#1''}%
  \providecommand \bibnamefont  [1]{#1}%
  \providecommand \bibfnamefont [1]{#1}%
  \providecommand \citenamefont [1]{#1}%
  \providecommand \href@noop [0]{\@secondoftwo}%
  \providecommand \href [0]{\begingroup \@sanitize@url \@href}%
  \providecommand \@href[1]{\@@startlink{#1}\@@href}%
  \providecommand \@@href[1]{\endgroup#1\@@endlink}%
  \providecommand \@sanitize@url [0]{\catcode `\\12\catcode `\$12\catcode
    `\&12\catcode `\#12\catcode `\^12\catcode `\_12\catcode `\%12\relax}%
  \providecommand \@@startlink[1]{}%
  \providecommand \@@endlink[0]{}%
  \providecommand \url  [0]{\begingroup\@sanitize@url \@url }%
  \providecommand \@url [1]{\endgroup\@href {#1}{\urlprefix }}%
  \providecommand \urlprefix  [0]{URL }%
  \providecommand \Eprint [0]{\href }%
  \providecommand \doibase [0]{http://dx.doi.org/}%
  \providecommand \selectlanguage [0]{\@gobble}%
  \providecommand \bibinfo  [0]{\@secondoftwo}%
  \providecommand \bibfield  [0]{\@secondoftwo}%
  \providecommand \translation [1]{[#1]}%
  \providecommand \BibitemOpen [0]{}%
  \providecommand \bibitemStop [0]{}%
  \providecommand \bibitemNoStop [0]{.\EOS\space}%
  \providecommand \EOS [0]{\spacefactor3000\relax}%
  \providecommand \BibitemShut  [1]{\csname bibitem#1\endcsname}%
  \let\auto@bib@innerbib\@empty
  \bibitem [{\citenamefont {Sturchio}\ \emph {et~al.}(2004)\citenamefont
    {Sturchio}, \citenamefont {Du}, \citenamefont {Purtschert}, \citenamefont
    {Lehmann}, \citenamefont {Sultan}, \citenamefont {Patterson}, \citenamefont
    {Lu}, \citenamefont {Mueller}, \citenamefont {Bigler}, \citenamefont
    {Bailey}, \citenamefont {O'Connor}, \citenamefont {Young}, \citenamefont
    {Lorenzo}, \citenamefont {Becker}, \citenamefont {El~Alfy}, \citenamefont
    {El~Kaliouby}, \citenamefont {Dawood},\ and\ \citenamefont
    {Abdallah}}]{grl04}%
    \BibitemOpen
    \bibfield  {author} {\bibinfo {author} {\bibfnamefont {N.~C.}\ \bibnamefont
    {Sturchio}}, \bibinfo {author} {\bibfnamefont {X.}~\bibnamefont {Du}},
    \bibinfo {author} {\bibfnamefont {R.}~\bibnamefont {Purtschert}}, \bibinfo
    {author} {\bibfnamefont {B.~E.}\ \bibnamefont {Lehmann}}, \bibinfo {author}
    {\bibfnamefont {M.}~\bibnamefont {Sultan}}, \bibinfo {author} {\bibfnamefont
    {L.~J.}\ \bibnamefont {Patterson}}, \bibinfo {author} {\bibfnamefont {Z.-T.}\
    \bibnamefont {Lu}}, \bibinfo {author} {\bibfnamefont {P.}~\bibnamefont
    {Mueller}}, \bibinfo {author} {\bibfnamefont {T.}~\bibnamefont {Bigler}},
    \bibinfo {author} {\bibfnamefont {K.}~\bibnamefont {Bailey}}, \bibinfo
    {author} {\bibfnamefont {T.~P.}\ \bibnamefont {O'Connor}}, \bibinfo {author}
    {\bibfnamefont {L.}~\bibnamefont {Young}}, \bibinfo {author} {\bibfnamefont
    {R.}~\bibnamefont {Lorenzo}}, \bibinfo {author} {\bibfnamefont
    {R.}~\bibnamefont {Becker}}, \bibinfo {author} {\bibfnamefont
    {Z.}~\bibnamefont {El~Alfy}}, \bibinfo {author} {\bibfnamefont
    {B.}~\bibnamefont {El~Kaliouby}}, \bibinfo {author} {\bibfnamefont
    {Y.}~\bibnamefont {Dawood}}, \ and\ \bibinfo {author} {\bibfnamefont
    {A.~M.~A.}\ \bibnamefont {Abdallah}},\ }\href {\doibase 10.1029/2003GL019234}
    {\bibfield  {journal} {\bibinfo  {journal} {Geophysical Research Letters}\
    }\textbf {\bibinfo {volume} {31}} (\bibinfo {year} {2004}),\
    10.1029/2003GL019234}\BibitemShut {NoStop}%
  \bibitem [{\citenamefont {Jiang}\ \emph {et~al.}(2011)\citenamefont {Jiang},
    \citenamefont {Williams}, \citenamefont {Bailey}, \citenamefont {Davis},
    \citenamefont {Hu}, \citenamefont {Lu}, \citenamefont {O'Connor},
    \citenamefont {Purtschert}, \citenamefont {Sturchio}, \citenamefont {Sun},\
    and\ \citenamefont {Mueller}}]{ar39prl}%
    \BibitemOpen
    \bibfield  {author} {\bibinfo {author} {\bibfnamefont {W.}~\bibnamefont
    {Jiang}}, \bibinfo {author} {\bibfnamefont {W.}~\bibnamefont {Williams}},
    \bibinfo {author} {\bibfnamefont {K.}~\bibnamefont {Bailey}}, \bibinfo
    {author} {\bibfnamefont {A.~M.}\ \bibnamefont {Davis}}, \bibinfo {author}
    {\bibfnamefont {S.-M.}\ \bibnamefont {Hu}}, \bibinfo {author} {\bibfnamefont
    {Z.-T.}\ \bibnamefont {Lu}}, \bibinfo {author} {\bibfnamefont {T.~P.}\
    \bibnamefont {O'Connor}}, \bibinfo {author} {\bibfnamefont {R.}~\bibnamefont
    {Purtschert}}, \bibinfo {author} {\bibfnamefont {N.~C.}\ \bibnamefont
    {Sturchio}}, \bibinfo {author} {\bibfnamefont {Y.~R.}\ \bibnamefont {Sun}}, \
    and\ \bibinfo {author} {\bibfnamefont {P.}~\bibnamefont {Mueller}},\ }\href
    {\doibase 10.1103/PhysRevLett.106.103001} {\bibfield  {journal} {\bibinfo
    {journal} {Phys. Rev. Lett.}\ }\textbf {\bibinfo {volume} {106}},\ \bibinfo
    {pages} {103001} (\bibinfo {year} {2011})}\BibitemShut {NoStop}%
  \bibitem [{\citenamefont {Lu}\ \emph {et~al.}(2013)\citenamefont {Lu},
    \citenamefont {Mueller}, \citenamefont {Drake}, \citenamefont
    {N\"ortersh\"auser}, \citenamefont {Pieper},\ and\ \citenamefont
    {Yan}}]{Lu2013}%
    \BibitemOpen
    \bibfield  {author} {\bibinfo {author} {\bibfnamefont {Z.-T.}\ \bibnamefont
    {Lu}}, \bibinfo {author} {\bibfnamefont {P.}~\bibnamefont {Mueller}},
    \bibinfo {author} {\bibfnamefont {G.~W.~F.}\ \bibnamefont {Drake}}, \bibinfo
    {author} {\bibfnamefont {W.}~\bibnamefont {N\"ortersh\"auser}}, \bibinfo
    {author} {\bibfnamefont {S.~C.}\ \bibnamefont {Pieper}}, \ and\ \bibinfo
    {author} {\bibfnamefont {Z.-C.}\ \bibnamefont {Yan}},\ }\href {\doibase
    10.1103/RevModPhys.85.1383} {\bibfield  {journal} {\bibinfo  {journal} {Rev.
    Mod. Phys.}\ }\textbf {\bibinfo {volume} {85}},\ \bibinfo {pages} {1383}
    (\bibinfo {year} {2013})}\BibitemShut {NoStop}%
  \bibitem [{\citenamefont {Mueller}\ \emph {et~al.}(2007)\citenamefont
    {Mueller}, \citenamefont {Sulai}, \citenamefont {Villari}, \citenamefont
    {Alc\'antara-N\'u\~nez}, \citenamefont {Alves-Cond\'e}, \citenamefont
    {Bailey}, \citenamefont {Drake}, \citenamefont {Dubois}, \citenamefont
    {El\'eon}, \citenamefont {Gaubert}, \citenamefont {Holt}, \citenamefont
    {Janssens}, \citenamefont {Lecesne}, \citenamefont {Lu}, \citenamefont
    {O'Connor}, \citenamefont {Saint-Laurent}, \citenamefont {Thomas},\ and\
    \citenamefont {Wang}}]{Mueller2007}%
    \BibitemOpen
    \bibfield  {author} {\bibinfo {author} {\bibfnamefont {P.}~\bibnamefont
    {Mueller}}, \bibinfo {author} {\bibfnamefont {I.~A.}\ \bibnamefont {Sulai}},
    \bibinfo {author} {\bibfnamefont {A.~C.~C.}\ \bibnamefont {Villari}},
    \bibinfo {author} {\bibfnamefont {J.~A.}\ \bibnamefont
    {Alc\'antara-N\'u\~nez}}, \bibinfo {author} {\bibfnamefont {R.}~\bibnamefont
    {Alves-Cond\'e}}, \bibinfo {author} {\bibfnamefont {K.}~\bibnamefont
    {Bailey}}, \bibinfo {author} {\bibfnamefont {G.~W.~F.}\ \bibnamefont
    {Drake}}, \bibinfo {author} {\bibfnamefont {M.}~\bibnamefont {Dubois}},
    \bibinfo {author} {\bibfnamefont {C.}~\bibnamefont {El\'eon}}, \bibinfo
    {author} {\bibfnamefont {G.}~\bibnamefont {Gaubert}}, \bibinfo {author}
    {\bibfnamefont {R.~J.}\ \bibnamefont {Holt}}, \bibinfo {author}
    {\bibfnamefont {R.~V.~F.}\ \bibnamefont {Janssens}}, \bibinfo {author}
    {\bibfnamefont {N.}~\bibnamefont {Lecesne}}, \bibinfo {author} {\bibfnamefont
    {Z.-T.}\ \bibnamefont {Lu}}, \bibinfo {author} {\bibfnamefont {T.~P.}\
    \bibnamefont {O'Connor}}, \bibinfo {author} {\bibfnamefont {M.-G.}\
    \bibnamefont {Saint-Laurent}}, \bibinfo {author} {\bibfnamefont {J.-C.}\
    \bibnamefont {Thomas}}, \ and\ \bibinfo {author} {\bibfnamefont {L.-B.}\
    \bibnamefont {Wang}},\ }\href {\doibase 10.1103/PhysRevLett.99.252501}
    {\bibfield  {journal} {\bibinfo  {journal} {Phys. Rev. Lett.}\ }\textbf
    {\bibinfo {volume} {99}},\ \bibinfo {pages} {252501} (\bibinfo {year}
    {2007})}\BibitemShut {NoStop}%
  \bibitem [{\citenamefont {Wang}\ \emph {et~al.}(2004)\citenamefont {Wang},
    \citenamefont {Mueller}, \citenamefont {Bailey}, \citenamefont {Drake},
    \citenamefont {Greene}, \citenamefont {Henderson}, \citenamefont {Holt},
    \citenamefont {Janssens}, \citenamefont {Jiang}, \citenamefont {Lu},
    \citenamefont {O'Connor}, \citenamefont {Pardo}, \citenamefont {Rehm},
    \citenamefont {Schiffer},\ and\ \citenamefont {Tang}}]{Wang2004}%
    \BibitemOpen
    \bibfield  {author} {\bibinfo {author} {\bibfnamefont {L.-B.}\ \bibnamefont
    {Wang}}, \bibinfo {author} {\bibfnamefont {P.}~\bibnamefont {Mueller}},
    \bibinfo {author} {\bibfnamefont {K.}~\bibnamefont {Bailey}}, \bibinfo
    {author} {\bibfnamefont {G.~W.~F.}\ \bibnamefont {Drake}}, \bibinfo {author}
    {\bibfnamefont {J.~P.}\ \bibnamefont {Greene}}, \bibinfo {author}
    {\bibfnamefont {D.}~\bibnamefont {Henderson}}, \bibinfo {author}
    {\bibfnamefont {R.~J.}\ \bibnamefont {Holt}}, \bibinfo {author}
    {\bibfnamefont {R.~V.~F.}\ \bibnamefont {Janssens}}, \bibinfo {author}
    {\bibfnamefont {C.~L.}\ \bibnamefont {Jiang}}, \bibinfo {author}
    {\bibfnamefont {Z.-T.}\ \bibnamefont {Lu}}, \bibinfo {author} {\bibfnamefont
    {T.~P.}\ \bibnamefont {O'Connor}}, \bibinfo {author} {\bibfnamefont {R.~C.}\
    \bibnamefont {Pardo}}, \bibinfo {author} {\bibfnamefont {K.~E.}\ \bibnamefont
    {Rehm}}, \bibinfo {author} {\bibfnamefont {J.~P.}\ \bibnamefont {Schiffer}},
    \ and\ \bibinfo {author} {\bibfnamefont {X.~D.}\ \bibnamefont {Tang}},\
    }\href {\doibase 10.1103/PhysRevLett.93.142501} {\bibfield  {journal}
    {\bibinfo  {journal} {Phys. Rev. Lett.}\ }\textbf {\bibinfo {volume} {93}},\
    \bibinfo {pages} {142501} (\bibinfo {year} {2004})}\BibitemShut {NoStop}%
  \bibitem [{\citenamefont {Vetter}\ \emph {et~al.}(2008)\citenamefont {Vetter},
    \citenamefont {Abo-Shaeer}, \citenamefont {Freedman},\ and\ \citenamefont
    {Maruyama}}]{na21prc}%
    \BibitemOpen
    \bibfield  {author} {\bibinfo {author} {\bibfnamefont {P.~A.}\ \bibnamefont
    {Vetter}}, \bibinfo {author} {\bibfnamefont {J.~R.}\ \bibnamefont
    {Abo-Shaeer}}, \bibinfo {author} {\bibfnamefont {S.~J.}\ \bibnamefont
    {Freedman}}, \ and\ \bibinfo {author} {\bibfnamefont {R.}~\bibnamefont
    {Maruyama}},\ }\href {\doibase 10.1103/PhysRevC.77.035502} {\bibfield
    {journal} {\bibinfo  {journal} {Phys. Rev. C}\ }\textbf {\bibinfo {volume}
    {77}},\ \bibinfo {pages} {035502} (\bibinfo {year} {2008})}\BibitemShut
    {NoStop}%
  \bibitem [{\citenamefont {Fenker}\ \emph {et~al.}(2018)\citenamefont {Fenker},
    \citenamefont {Gorelov}, \citenamefont {Melconian}, \citenamefont {Behr},
    \citenamefont {Anholm}, \citenamefont {Ashery}, \citenamefont {Behling},
    \citenamefont {Cohen}, \citenamefont {Craiciu}, \citenamefont {Gwinner},
    \citenamefont {McNeil}, \citenamefont {Mehlman}, \citenamefont {Olchanski},
    \citenamefont {Shidling}, \citenamefont {Smale},\ and\ \citenamefont
    {Warner}}]{k37prl}%
    \BibitemOpen
    \bibfield  {author} {\bibinfo {author} {\bibfnamefont {B.}~\bibnamefont
    {Fenker}}, \bibinfo {author} {\bibfnamefont {A.}~\bibnamefont {Gorelov}},
    \bibinfo {author} {\bibfnamefont {D.}~\bibnamefont {Melconian}}, \bibinfo
    {author} {\bibfnamefont {J.~A.}\ \bibnamefont {Behr}}, \bibinfo {author}
    {\bibfnamefont {M.}~\bibnamefont {Anholm}}, \bibinfo {author} {\bibfnamefont
    {D.}~\bibnamefont {Ashery}}, \bibinfo {author} {\bibfnamefont {R.~S.}\
    \bibnamefont {Behling}}, \bibinfo {author} {\bibfnamefont {I.}~\bibnamefont
    {Cohen}}, \bibinfo {author} {\bibfnamefont {I.}~\bibnamefont {Craiciu}},
    \bibinfo {author} {\bibfnamefont {G.}~\bibnamefont {Gwinner}}, \bibinfo
    {author} {\bibfnamefont {J.}~\bibnamefont {McNeil}}, \bibinfo {author}
    {\bibfnamefont {M.}~\bibnamefont {Mehlman}}, \bibinfo {author} {\bibfnamefont
    {K.}~\bibnamefont {Olchanski}}, \bibinfo {author} {\bibfnamefont {P.~D.}\
    \bibnamefont {Shidling}}, \bibinfo {author} {\bibfnamefont {S.}~\bibnamefont
    {Smale}}, \ and\ \bibinfo {author} {\bibfnamefont {C.~L.}\ \bibnamefont
    {Warner}},\ }\href {\doibase 10.1103/PhysRevLett.120.062502} {\bibfield
    {journal} {\bibinfo  {journal} {Phys. Rev. Lett.}\ }\textbf {\bibinfo
    {volume} {120}},\ \bibinfo {pages} {062502} (\bibinfo {year}
    {2018})}\BibitemShut {NoStop}%
  \bibitem [{\citenamefont {González-Alonso}\ \emph {et~al.}(2019)\citenamefont
    {González-Alonso}, \citenamefont {Naviliat-Cuncic},\ and\ \citenamefont
    {Severijns}}]{ppnp19}%
    \BibitemOpen
    \bibfield  {author} {\bibinfo {author} {\bibfnamefont {M.}~\bibnamefont
    {González-Alonso}}, \bibinfo {author} {\bibfnamefont {O.}~\bibnamefont
    {Naviliat-Cuncic}}, \ and\ \bibinfo {author} {\bibfnamefont {N.}~\bibnamefont
    {Severijns}},\ }\href {\doibase https://doi.org/10.1016/j.ppnp.2018.08.002}
    {\bibfield  {journal} {\bibinfo  {journal} {Progress in Particle and Nuclear
    Physics}\ }\textbf {\bibinfo {volume} {104}},\ \bibinfo {pages} {165 }
    (\bibinfo {year} {2019})}\BibitemShut {NoStop}%
  \bibitem [{\citenamefont {{Tandecki}}\ \emph {et~al.}(2013)\citenamefont
    {{Tandecki}}, \citenamefont {{Zhang}}, \citenamefont {{Collister}},
    \citenamefont {{Aubin}}, \citenamefont {{Behr}}, \citenamefont {{Gomez}},
    \citenamefont {{Gwinner}}, \citenamefont {{Orozco}},\ and\ \citenamefont
    {{Pearson}}}]{Tandecki2006}%
    \BibitemOpen
    \bibfield  {author} {\bibinfo {author} {\bibfnamefont {M.}~\bibnamefont
    {{Tandecki}}}, \bibinfo {author} {\bibfnamefont {J.}~\bibnamefont {{Zhang}}},
    \bibinfo {author} {\bibfnamefont {R.}~\bibnamefont {{Collister}}}, \bibinfo
    {author} {\bibfnamefont {S.}~\bibnamefont {{Aubin}}}, \bibinfo {author}
    {\bibfnamefont {J.~A.}\ \bibnamefont {{Behr}}}, \bibinfo {author}
    {\bibfnamefont {E.}~\bibnamefont {{Gomez}}}, \bibinfo {author} {\bibfnamefont
    {G.}~\bibnamefont {{Gwinner}}}, \bibinfo {author} {\bibfnamefont {L.~A.}\
    \bibnamefont {{Orozco}}}, \ and\ \bibinfo {author} {\bibfnamefont {M.~R.}\
    \bibnamefont {{Pearson}}},\ }\href {\doibase 10.1088/1748-0221/8/12/P12006}
    {\bibfield  {journal} {\bibinfo  {journal} {J. Instrum.}\ }\textbf {\bibinfo
    {volume} {8}},\ \bibinfo {eid} {P12006} (\bibinfo {year} {2013})},\ \Eprint
    {http://arxiv.org/abs/1312.3562} {arXiv:1312.3562 [physics.atom-ph]}
    \BibitemShut {NoStop}%
  \bibitem [{\citenamefont {Parker}\ \emph {et~al.}(2012)\citenamefont {Parker},
    \citenamefont {Dietrich}, \citenamefont {Bailey}, \citenamefont {Greene},
    \citenamefont {Holt}, \citenamefont {Kalita}, \citenamefont {Korsch},
    \citenamefont {Lu}, \citenamefont {Mueller}, \citenamefont {O'Connor},
    \citenamefont {Singh}, \citenamefont {Sulai},\ and\ \citenamefont
    {Trimble}}]{Parker2012}%
    \BibitemOpen
    \bibfield  {author} {\bibinfo {author} {\bibfnamefont {R.~H.}\ \bibnamefont
    {Parker}}, \bibinfo {author} {\bibfnamefont {M.~R.}\ \bibnamefont
    {Dietrich}}, \bibinfo {author} {\bibfnamefont {K.}~\bibnamefont {Bailey}},
    \bibinfo {author} {\bibfnamefont {J.~P.}\ \bibnamefont {Greene}}, \bibinfo
    {author} {\bibfnamefont {R.~J.}\ \bibnamefont {Holt}}, \bibinfo {author}
    {\bibfnamefont {M.~R.}\ \bibnamefont {Kalita}}, \bibinfo {author}
    {\bibfnamefont {W.}~\bibnamefont {Korsch}}, \bibinfo {author} {\bibfnamefont
    {Z.-T.}\ \bibnamefont {Lu}}, \bibinfo {author} {\bibfnamefont
    {P.}~\bibnamefont {Mueller}}, \bibinfo {author} {\bibfnamefont {T.~P.}\
    \bibnamefont {O'Connor}}, \bibinfo {author} {\bibfnamefont {J.}~\bibnamefont
    {Singh}}, \bibinfo {author} {\bibfnamefont {I.~A.}\ \bibnamefont {Sulai}}, \
    and\ \bibinfo {author} {\bibfnamefont {W.~L.}\ \bibnamefont {Trimble}},\
    }\href {\doibase 10.1103/PhysRevC.86.065503} {\bibfield  {journal} {\bibinfo
    {journal} {Phys. Rev. C}\ }\textbf {\bibinfo {volume} {86}},\ \bibinfo
    {pages} {065503} (\bibinfo {year} {2012})}\BibitemShut {NoStop}%
  \bibitem [{\citenamefont {Chupp}\ \emph {et~al.}(2019)\citenamefont {Chupp},
    \citenamefont {Fierlinger}, \citenamefont {Ramsey-Musolf},\ and\
    \citenamefont {Singh}}]{rmp2019}%
    \BibitemOpen
    \bibfield  {author} {\bibinfo {author} {\bibfnamefont {T.~E.}\ \bibnamefont
    {Chupp}}, \bibinfo {author} {\bibfnamefont {P.}~\bibnamefont {Fierlinger}},
    \bibinfo {author} {\bibfnamefont {M.~J.}\ \bibnamefont {Ramsey-Musolf}}, \
    and\ \bibinfo {author} {\bibfnamefont {J.~T.}\ \bibnamefont {Singh}},\
    }\href@noop {} {\bibfield  {journal} {\bibinfo  {journal} {Rev. Mod. Phys.}\
    }\textbf {\bibinfo {volume} {91}},\ \bibinfo {pages} {015001} (\bibinfo
    {year} {2019})}\BibitemShut {NoStop}%
  \bibitem [{\citenamefont {Auerbach}\ \emph {et~al.}(1996)\citenamefont
    {Auerbach}, \citenamefont {Flambaum},\ and\ \citenamefont
    {Spevak}}]{Auerbach1996}%
    \BibitemOpen
    \bibfield  {author} {\bibinfo {author} {\bibfnamefont {N.}~\bibnamefont
    {Auerbach}}, \bibinfo {author} {\bibfnamefont {V.~V.}\ \bibnamefont
    {Flambaum}}, \ and\ \bibinfo {author} {\bibfnamefont {V.}~\bibnamefont
    {Spevak}},\ }\href {\doibase 10.1103/PhysRevLett.76.4316} {\bibfield
    {journal} {\bibinfo  {journal} {Phys. Rev. Lett.}\ }\textbf {\bibinfo
    {volume} {76}},\ \bibinfo {pages} {4316} (\bibinfo {year}
    {1996})}\BibitemShut {NoStop}%
  \bibitem [{\citenamefont {Dobaczewski}\ and\ \citenamefont
    {Engel}(2005)}]{Dobaczewski2005}%
    \BibitemOpen
    \bibfield  {author} {\bibinfo {author} {\bibfnamefont {J.}~\bibnamefont
    {Dobaczewski}}\ and\ \bibinfo {author} {\bibfnamefont {J.}~\bibnamefont
    {Engel}},\ }\href {\doibase 10.1103/PhysRevLett.94.232502} {\bibfield
    {journal} {\bibinfo  {journal} {Phys. Rev. Lett.}\ }\textbf {\bibinfo
    {volume} {94}},\ \bibinfo {pages} {232502} (\bibinfo {year}
    {2005})}\BibitemShut {NoStop}%
  \bibitem [{\citenamefont {Bishof}\ \emph {et~al.}(2016)\citenamefont {Bishof},
    \citenamefont {Parker}, \citenamefont {Bailey}, \citenamefont {Greene},
    \citenamefont {Holt}, \citenamefont {Kalita}, \citenamefont {Korsch},
    \citenamefont {Lemke}, \citenamefont {Lu}, \citenamefont {Mueller},
    \citenamefont {O'Connor}, \citenamefont {Singh},\ and\ \citenamefont
    {Dietrich}}]{Bishof2016}%
    \BibitemOpen
    \bibfield  {author} {\bibinfo {author} {\bibfnamefont {M.}~\bibnamefont
    {Bishof}}, \bibinfo {author} {\bibfnamefont {R.~H.}\ \bibnamefont {Parker}},
    \bibinfo {author} {\bibfnamefont {K.~G.}\ \bibnamefont {Bailey}}, \bibinfo
    {author} {\bibfnamefont {J.~P.}\ \bibnamefont {Greene}}, \bibinfo {author}
    {\bibfnamefont {R.~J.}\ \bibnamefont {Holt}}, \bibinfo {author}
    {\bibfnamefont {M.~R.}\ \bibnamefont {Kalita}}, \bibinfo {author}
    {\bibfnamefont {W.}~\bibnamefont {Korsch}}, \bibinfo {author} {\bibfnamefont
    {N.~D.}\ \bibnamefont {Lemke}}, \bibinfo {author} {\bibfnamefont {Z.-T.}\
    \bibnamefont {Lu}}, \bibinfo {author} {\bibfnamefont {P.}~\bibnamefont
    {Mueller}}, \bibinfo {author} {\bibfnamefont {T.~P.}\ \bibnamefont
    {O'Connor}}, \bibinfo {author} {\bibfnamefont {J.~T.}\ \bibnamefont {Singh}},
    \ and\ \bibinfo {author} {\bibfnamefont {M.~R.}\ \bibnamefont {Dietrich}},\
    }\href {\doibase 10.1103/PhysRevC.94.025501} {\bibfield  {journal} {\bibinfo
    {journal} {Phys. Rev. C}\ }\textbf {\bibinfo {volume} {94}},\ \bibinfo
    {pages} {025501} (\bibinfo {year} {2016})}\BibitemShut {NoStop}%
  \bibitem [{\citenamefont {Blatt}\ \emph {et~al.}(1986)\citenamefont {Blatt},
    \citenamefont {Ertmer}, \citenamefont {Zoller},\ and\ \citenamefont
    {Hall}}]{Blatt1986}%
    \BibitemOpen
    \bibfield  {author} {\bibinfo {author} {\bibfnamefont {R.}~\bibnamefont
    {Blatt}}, \bibinfo {author} {\bibfnamefont {W.}~\bibnamefont {Ertmer}},
    \bibinfo {author} {\bibfnamefont {P.}~\bibnamefont {Zoller}}, \ and\ \bibinfo
    {author} {\bibfnamefont {J.~L.}\ \bibnamefont {Hall}},\ }\href {\doibase
    10.1103/PhysRevA.34.3022} {\bibfield  {journal} {\bibinfo  {journal} {Phys.
    Rev. A}\ }\textbf {\bibinfo {volume} {34}},\ \bibinfo {pages} {3022}
    (\bibinfo {year} {1986})}\BibitemShut {NoStop}%
  \bibitem [{\citenamefont {Javanainen}(1992)}]{Javanainen1992}%
    \BibitemOpen
    \bibfield  {author} {\bibinfo {author} {\bibfnamefont {J.}~\bibnamefont
    {Javanainen}},\ }\href@noop {} {\bibfield  {journal} {\bibinfo  {journal}
    {Phys. Rev. A}\ }\textbf {\bibinfo {volume} {46}},\ \bibinfo {pages} {5819}
    (\bibinfo {year} {1992})}\BibitemShut {NoStop}%
  \bibitem [{\citenamefont {Vredenbregt}\ and\ \citenamefont {van
    Leeuwen}(2003)}]{Vredenbregt2003}%
    \BibitemOpen
    \bibfield  {author} {\bibinfo {author} {\bibfnamefont {E.~J.~D.}\
    \bibnamefont {Vredenbregt}}\ and\ \bibinfo {author} {\bibfnamefont
    {K.~A.~H.}\ \bibnamefont {van Leeuwen}},\ }\href {\doibase 10.1119/1.1578063}
    {\bibfield  {journal} {\bibinfo  {journal} {Am. J. Phys}\ }\textbf {\bibinfo
    {volume} {71}},\ \bibinfo {pages} {760} (\bibinfo {year} {2003})}\BibitemShut
    {NoStop}%
  \bibitem [{\citenamefont {Hamamda}\ \emph {et~al.}(2015)\citenamefont
    {Hamamda}, \citenamefont {Taillandier-Loize}, \citenamefont {Baudon},
    \citenamefont {Dutier}, \citenamefont {Perales},\ and\ \citenamefont
    {Ducloy}}]{Hamamda2015}%
    \BibitemOpen
    \bibfield  {author} {\bibinfo {author} {\bibfnamefont {M.}~\bibnamefont
    {Hamamda}}, \bibinfo {author} {\bibfnamefont {T.}~\bibnamefont
    {Taillandier-Loize}}, \bibinfo {author} {\bibfnamefont {J.}~\bibnamefont
    {Baudon}}, \bibinfo {author} {\bibfnamefont {G.}~\bibnamefont {Dutier}},
    \bibinfo {author} {\bibfnamefont {F.}~\bibnamefont {Perales}}, \ and\
    \bibinfo {author} {\bibfnamefont {M.}~\bibnamefont {Ducloy}},\ }\href@noop {}
    {\bibfield  {journal} {\bibinfo  {journal} {Eur. Phys. J. Appl. Phys.}\
    }\textbf {\bibinfo {volume} {71}},\ \bibinfo {pages} {30502} (\bibinfo {year}
    {2015})}\BibitemShut {NoStop}%
  \bibitem [{\citenamefont {Castin}\ and\ \citenamefont
    {M{\o}lmer}(1995)}]{Castin1995}%
    \BibitemOpen
    \bibfield  {author} {\bibinfo {author} {\bibfnamefont {Y.}~\bibnamefont
    {Castin}}\ and\ \bibinfo {author} {\bibfnamefont {K.}~\bibnamefont
    {M{\o}lmer}},\ }\href@noop {} {\bibfield  {journal} {\bibinfo  {journal}
    {Phys Rev Lett}\ }\textbf {\bibinfo {volume} {74}},\ \bibinfo {pages} {3772}
    (\bibinfo {year} {1995})}\BibitemShut {NoStop}%
  \bibitem [{\citenamefont {Dunn}\ and\ \citenamefont {Greene}(2006)}]{Dunn2006}%
    \BibitemOpen
    \bibfield  {author} {\bibinfo {author} {\bibfnamefont {J.~W.}\ \bibnamefont
    {Dunn}}\ and\ \bibinfo {author} {\bibfnamefont {C.~H.}\ \bibnamefont
    {Greene}},\ }\href@noop {} {\bibfield  {journal} {\bibinfo  {journal} {Phys.
    Rev. A}\ }\textbf {\bibinfo {volume} {73}},\ \bibinfo {pages} {033421}
    (\bibinfo {year} {2006})}\BibitemShut {NoStop}%
  \bibitem [{\citenamefont {Griffith}\ \emph {et~al.}(2009)\citenamefont
    {Griffith}, \citenamefont {Swallows}, \citenamefont {Loftus}, \citenamefont
    {Romalis}, \citenamefont {Heckel},\ and\ \citenamefont
    {Fortson}}]{Griffith2009}%
    \BibitemOpen
    \bibfield  {author} {\bibinfo {author} {\bibfnamefont {W.~C.}\ \bibnamefont
    {Griffith}}, \bibinfo {author} {\bibfnamefont {M.~D.}\ \bibnamefont
    {Swallows}}, \bibinfo {author} {\bibfnamefont {T.~H.}\ \bibnamefont
    {Loftus}}, \bibinfo {author} {\bibfnamefont {M.~V.}\ \bibnamefont {Romalis}},
    \bibinfo {author} {\bibfnamefont {B.~R.}\ \bibnamefont {Heckel}}, \ and\
    \bibinfo {author} {\bibfnamefont {E.~N.}\ \bibnamefont {Fortson}},\ }\href
    {\doibase 10.1103/PhysRevLett.102.101601} {\bibfield  {journal} {\bibinfo
    {journal} {Phys. Rev. Lett.}\ }\textbf {\bibinfo {volume} {102}},\ \bibinfo
    {pages} {101601} (\bibinfo {year} {2009})}\BibitemShut {NoStop}%
  \bibitem [{\citenamefont {Holt}\ \emph {et~al.}(2010)\citenamefont {Holt},
    \citenamefont {Ahmad}, \citenamefont {Bailey}, \citenamefont {Graner},
    \citenamefont {Greene} \emph {et~al.}}]{Holt2010}%
    \BibitemOpen
    \bibfield  {author} {\bibinfo {author} {\bibfnamefont {R.}~\bibnamefont
    {Holt}}, \bibinfo {author} {\bibfnamefont {I.}~\bibnamefont {Ahmad}},
    \bibinfo {author} {\bibfnamefont {K.}~\bibnamefont {Bailey}}, \bibinfo
    {author} {\bibfnamefont {B.}~\bibnamefont {Graner}}, \bibinfo {author}
    {\bibfnamefont {J.}~\bibnamefont {Greene}},  \emph {et~al.},\ }\href
    {\doibase 10.1016/j.nuclphysa.2010.05.013} {\bibfield  {journal} {\bibinfo
    {journal} {Nucl.Phys.}\ }\textbf {\bibinfo {volume} {A844}},\ \bibinfo
    {pages} {53c} (\bibinfo {year} {2010})}\BibitemShut {NoStop}%
  \bibitem [{\citenamefont {Guest}\ \emph {et~al.}(2007)\citenamefont {Guest},
    \citenamefont {Scielzo}, \citenamefont {Ahmad}, \citenamefont {Bailey},
    \citenamefont {Greene}, \citenamefont {Holt}, \citenamefont {Lu},
    \citenamefont {O'Connor},\ and\ \citenamefont {Potterveld}}]{Guest2007}%
    \BibitemOpen
    \bibfield  {author} {\bibinfo {author} {\bibfnamefont {J.~R.}\ \bibnamefont
    {Guest}}, \bibinfo {author} {\bibfnamefont {N.~D.}\ \bibnamefont {Scielzo}},
    \bibinfo {author} {\bibfnamefont {I.}~\bibnamefont {Ahmad}}, \bibinfo
    {author} {\bibfnamefont {K.}~\bibnamefont {Bailey}}, \bibinfo {author}
    {\bibfnamefont {J.~P.}\ \bibnamefont {Greene}}, \bibinfo {author}
    {\bibfnamefont {R.~J.}\ \bibnamefont {Holt}}, \bibinfo {author}
    {\bibfnamefont {Z.-T.}\ \bibnamefont {Lu}}, \bibinfo {author} {\bibfnamefont
    {T.~P.}\ \bibnamefont {O'Connor}}, \ and\ \bibinfo {author} {\bibfnamefont
    {D.~H.}\ \bibnamefont {Potterveld}},\ }\href {\doibase
    10.1103/PhysRevLett.98.093001} {\bibfield  {journal} {\bibinfo  {journal}
    {Phys. Rev. Lett.}\ }\textbf {\bibinfo {volume} {98}},\ \bibinfo {pages}
    {093001} (\bibinfo {year} {2007})}\BibitemShut {NoStop}%
  \bibitem [{\citenamefont {Parker}\ \emph {et~al.}(2015)\citenamefont {Parker},
    \citenamefont {Dietrich}, \citenamefont {Kalita}, \citenamefont {Lemke},
    \citenamefont {Bailey}, \citenamefont {Bishof}, \citenamefont {Greene},
    \citenamefont {Holt}, \citenamefont {Korsch}, \citenamefont {Lu},
    \citenamefont {Mueller}, \citenamefont {O'Connor},\ and\ \citenamefont
    {Singh}}]{Parker2015}%
    \BibitemOpen
    \bibfield  {author} {\bibinfo {author} {\bibfnamefont {R.~H.}\ \bibnamefont
    {Parker}}, \bibinfo {author} {\bibfnamefont {M.~R.}\ \bibnamefont
    {Dietrich}}, \bibinfo {author} {\bibfnamefont {M.~R.}\ \bibnamefont
    {Kalita}}, \bibinfo {author} {\bibfnamefont {N.~D.}\ \bibnamefont {Lemke}},
    \bibinfo {author} {\bibfnamefont {K.~G.}\ \bibnamefont {Bailey}}, \bibinfo
    {author} {\bibfnamefont {M.}~\bibnamefont {Bishof}}, \bibinfo {author}
    {\bibfnamefont {J.~P.}\ \bibnamefont {Greene}}, \bibinfo {author}
    {\bibfnamefont {R.~J.}\ \bibnamefont {Holt}}, \bibinfo {author}
    {\bibfnamefont {W.}~\bibnamefont {Korsch}}, \bibinfo {author} {\bibfnamefont
    {Z.-T.}\ \bibnamefont {Lu}}, \bibinfo {author} {\bibfnamefont
    {P.}~\bibnamefont {Mueller}}, \bibinfo {author} {\bibfnamefont {T.~P.}\
    \bibnamefont {O'Connor}}, \ and\ \bibinfo {author} {\bibfnamefont {J.~T.}\
    \bibnamefont {Singh}},\ }\href {\doibase 10.1103/PhysRevLett.114.233002}
    {\bibfield  {journal} {\bibinfo  {journal} {Phys. Rev. Lett.}\ }\textbf
    {\bibinfo {volume} {114}},\ \bibinfo {pages} {233002} (\bibinfo {year}
    {2015})}\BibitemShut {NoStop}%
  \bibitem [{\citenamefont {Breit}\ and\ \citenamefont
    {Wills}(1933)}]{Breit1933}%
    \BibitemOpen
    \bibfield  {author} {\bibinfo {author} {\bibfnamefont {G.}~\bibnamefont
    {Breit}}\ and\ \bibinfo {author} {\bibfnamefont {L.~A.}\ \bibnamefont
    {Wills}},\ }\href {\doibase 10.1103/PhysRev.44.470} {\bibfield  {journal}
    {\bibinfo  {journal} {Phys. Rev.}\ }\textbf {\bibinfo {volume} {44}},\
    \bibinfo {pages} {470} (\bibinfo {year} {1933})}\BibitemShut {NoStop}%
  \bibitem [{\citenamefont {Scielzo}\ \emph {et~al.}(2006)\citenamefont
    {Scielzo}, \citenamefont {Guest}, \citenamefont {Schulte}, \citenamefont
    {Ahmad}, \citenamefont {Bailey}, \citenamefont {Bowers}, \citenamefont
    {Holt}, \citenamefont {Lu}, \citenamefont {O'Connor},\ and\ \citenamefont
    {Potterveld}}]{Scielzo2006}%
    \BibitemOpen
    \bibfield  {author} {\bibinfo {author} {\bibfnamefont {N.~D.}\ \bibnamefont
    {Scielzo}}, \bibinfo {author} {\bibfnamefont {J.~R.}\ \bibnamefont {Guest}},
    \bibinfo {author} {\bibfnamefont {E.~C.}\ \bibnamefont {Schulte}}, \bibinfo
    {author} {\bibfnamefont {I.}~\bibnamefont {Ahmad}}, \bibinfo {author}
    {\bibfnamefont {K.}~\bibnamefont {Bailey}}, \bibinfo {author} {\bibfnamefont
    {D.~L.}\ \bibnamefont {Bowers}}, \bibinfo {author} {\bibfnamefont {R.~J.}\
    \bibnamefont {Holt}}, \bibinfo {author} {\bibfnamefont {Z.-T.}\ \bibnamefont
    {Lu}}, \bibinfo {author} {\bibfnamefont {T.~P.}\ \bibnamefont {O'Connor}}, \
    and\ \bibinfo {author} {\bibfnamefont {D.~H.}\ \bibnamefont {Potterveld}},\
    }\href {\doibase 10.1103/PhysRevA.73.010501} {\bibfield  {journal} {\bibinfo
    {journal} {Phys. Rev. A}\ }\textbf {\bibinfo {volume} {73}},\ \bibinfo
    {pages} {010501(R)} (\bibinfo {year} {2006})}\BibitemShut {NoStop}%
  \bibitem [{\citenamefont {Ludlow}\ \emph {et~al.}(2008)\citenamefont {Ludlow},
    \citenamefont {Zelevinsky}, \citenamefont {Campbell}, \citenamefont {Blatt},
    \citenamefont {Boyd}, \citenamefont {de~Miranda}, \citenamefont {Martin},
    \citenamefont {Thomsen}, \citenamefont {Foreman}, \citenamefont {Ye},
    \citenamefont {Fortier}, \citenamefont {Stalnaker}, \citenamefont {Diddams},
    \citenamefont {Le~Coq}, \citenamefont {Barber}, \citenamefont {Poli},
    \citenamefont {Lemke}, \citenamefont {Beck},\ and\ \citenamefont
    {Oates}}]{Ludlow2008}%
    \BibitemOpen
    \bibfield  {author} {\bibinfo {author} {\bibfnamefont {A.~D.}\ \bibnamefont
    {Ludlow}}, \bibinfo {author} {\bibfnamefont {T.}~\bibnamefont {Zelevinsky}},
    \bibinfo {author} {\bibfnamefont {G.~K.}\ \bibnamefont {Campbell}}, \bibinfo
    {author} {\bibfnamefont {S.}~\bibnamefont {Blatt}}, \bibinfo {author}
    {\bibfnamefont {M.~M.}\ \bibnamefont {Boyd}}, \bibinfo {author}
    {\bibfnamefont {M.~H.~G.}\ \bibnamefont {de~Miranda}}, \bibinfo {author}
    {\bibfnamefont {M.~J.}\ \bibnamefont {Martin}}, \bibinfo {author}
    {\bibfnamefont {J.~W.}\ \bibnamefont {Thomsen}}, \bibinfo {author}
    {\bibfnamefont {S.~M.}\ \bibnamefont {Foreman}}, \bibinfo {author}
    {\bibfnamefont {J.}~\bibnamefont {Ye}}, \bibinfo {author} {\bibfnamefont
    {T.~M.}\ \bibnamefont {Fortier}}, \bibinfo {author} {\bibfnamefont {J.~E.}\
    \bibnamefont {Stalnaker}}, \bibinfo {author} {\bibfnamefont {S.~A.}\
    \bibnamefont {Diddams}}, \bibinfo {author} {\bibfnamefont {Y.}~\bibnamefont
    {Le~Coq}}, \bibinfo {author} {\bibfnamefont {Z.~W.}\ \bibnamefont {Barber}},
    \bibinfo {author} {\bibfnamefont {N.}~\bibnamefont {Poli}}, \bibinfo {author}
    {\bibfnamefont {N.~D.}\ \bibnamefont {Lemke}}, \bibinfo {author}
    {\bibfnamefont {K.~M.}\ \bibnamefont {Beck}}, \ and\ \bibinfo {author}
    {\bibfnamefont {C.~W.}\ \bibnamefont {Oates}},\ }\href@noop {} {\bibfield
    {journal} {\bibinfo  {journal} {Science}\ }\textbf {\bibinfo {volume}
    {319}},\ \bibinfo {pages} {1805} (\bibinfo {year} {2008})}\BibitemShut
    {NoStop}%
  \bibitem [{\citenamefont {Loftus}\ \emph {et~al.}(2004)\citenamefont {Loftus},
    \citenamefont {Ido}, \citenamefont {Boyd}, \citenamefont {Ludlow},\ and\
    \citenamefont {Ye}}]{Loftus2004}%
    \BibitemOpen
    \bibfield  {author} {\bibinfo {author} {\bibfnamefont {T.~H.}\ \bibnamefont
    {Loftus}}, \bibinfo {author} {\bibfnamefont {T.}~\bibnamefont {Ido}},
    \bibinfo {author} {\bibfnamefont {M.~M.}\ \bibnamefont {Boyd}}, \bibinfo
    {author} {\bibfnamefont {A.~D.}\ \bibnamefont {Ludlow}}, \ and\ \bibinfo
    {author} {\bibfnamefont {J.}~\bibnamefont {Ye}},\ }\href {\doibase
    10.1103/PhysRevA.70.063413} {\bibfield  {journal} {\bibinfo  {journal} {Phys.
    Rev. A}\ }\textbf {\bibinfo {volume} {70}},\ \bibinfo {pages} {063413}
    (\bibinfo {year} {2004})}\BibitemShut {NoStop}%
  \bibitem [{\citenamefont {Ludlow}(2008)}]{Ludlow2008a}%
    \BibitemOpen
    \bibfield  {author} {\bibinfo {author} {\bibfnamefont {A.~D.}\ \bibnamefont
    {Ludlow}},\ }\emph {\bibinfo {title} {The strontium optical lattice clock:
    optical spectroscopy with sub-hertz accuracy}},\ \href@noop {} {Ph.D.
    thesis},\ \bibinfo  {school} {University of Colorado-Boulder} (\bibinfo
    {year} {2008})\BibitemShut {NoStop}%
  \bibitem [{\citenamefont {Boyd}(2007)}]{Boyd2007}%
    \BibitemOpen
    \bibfield  {author} {\bibinfo {author} {\bibfnamefont {M.~M.}\ \bibnamefont
    {Boyd}},\ }\emph {\bibinfo {title} {High Precision Spectroscopy of Strontium
    in an Optical Lattice: Towards a New Standard for Frequency and Time}},\
    \href@noop {} {Ph.D. thesis},\ \bibinfo  {school} {University of
    Colorado-Boulder} (\bibinfo {year} {2007})\BibitemShut {NoStop}%
  \bibitem [{\citenamefont {Martin}(2013)}]{Martin2013}%
    \BibitemOpen
    \bibfield  {author} {\bibinfo {author} {\bibfnamefont {M.~J.}\ \bibnamefont
    {Martin}},\ }\emph {\bibinfo {title} {Quantum Metrology and Many-Body
    Physics: Pushing the Frontier of the Optical Lattice Clock}},\ \href@noop {}
    {Ph.D. thesis},\ \bibinfo  {school} {University of Colorado-Boulder}
    (\bibinfo {year} {2013})\BibitemShut {NoStop}%
  \bibitem [{\citenamefont {Barrett}\ \emph {et~al.}(1991)\citenamefont
    {Barrett}, \citenamefont {Dapore-Schwartz}, \citenamefont {Ray},\ and\
    \citenamefont {Lafyatis}}]{Barrett1991}%
    \BibitemOpen
    \bibfield  {author} {\bibinfo {author} {\bibfnamefont {T.~E.}\ \bibnamefont
    {Barrett}}, \bibinfo {author} {\bibfnamefont {S.~W.}\ \bibnamefont
    {Dapore-Schwartz}}, \bibinfo {author} {\bibfnamefont {M.~D.}\ \bibnamefont
    {Ray}}, \ and\ \bibinfo {author} {\bibfnamefont {G.~P.}\ \bibnamefont
    {Lafyatis}},\ }\href {\doibase 10.1103/PhysRevLett.67.3483} {\bibfield
    {journal} {\bibinfo  {journal} {Phys. Rev. Lett.}\ }\textbf {\bibinfo
    {volume} {67}},\ \bibinfo {pages} {3483} (\bibinfo {year}
    {1991})}\BibitemShut {NoStop}%
  \bibitem [{\citenamefont {Ludlow}(2013)}]{Ludlow2013}%
    \BibitemOpen
    \bibfield  {author} {\bibinfo {author} {\bibfnamefont {A.}~\bibnamefont
    {Ludlow}},\ }\href@noop {} {\enquote {\bibinfo {title} {Private
    communication},}\ } (\bibinfo {year} {2013})\BibitemShut {NoStop}%
  \bibitem [{\citenamefont {Venkattraman}\ and\ \citenamefont
    {Alexeenko}(2012)}]{Venka2012}%
    \BibitemOpen
    \bibfield  {author} {\bibinfo {author} {\bibfnamefont {A.}~\bibnamefont
    {Venkattraman}}\ and\ \bibinfo {author} {\bibfnamefont {A.~A.}\ \bibnamefont
    {Alexeenko}},\ }\href {\doibase 10.1063/1.3682375} {\bibfield  {journal}
    {\bibinfo  {journal} {Phys. Fluids}\ }\textbf {\bibinfo {volume} {24}},\
    \bibinfo {pages} {027101} (\bibinfo {year} {2012})}\BibitemShut {NoStop}%
  \bibitem [{\citenamefont {Library}\ \emph {et~al.}(1993)\citenamefont
    {Library}, \citenamefont {K{\"o}lbig}, \citenamefont {for Nuclear
    Research.~Computing},\ and\ \citenamefont {Group}}]{cernlib}%
    \BibitemOpen
    \bibfield  {author} {\bibinfo {author} {\bibfnamefont {C.~P.}\ \bibnamefont
    {Library}}, \bibinfo {author} {\bibfnamefont {K.}~\bibnamefont {K{\"o}lbig}},
    \bibinfo {author} {\bibfnamefont {E.~O.}\ \bibnamefont {for Nuclear
    Research.~Computing}}, \ and\ \bibinfo {author} {\bibfnamefont {N.~D. A.~S.}\
    \bibnamefont {Group}},\ }\href@noop {} {\emph {\bibinfo {title} {CERNLIB:
    Short Writeups}}}\ (\bibinfo  {publisher} {CERN},\ \bibinfo {year}
    {1993})\BibitemShut {NoStop}%
  \bibitem [{\citenamefont {Bogner}\ \emph {et~al.}(2013)\citenamefont {Bogner},
    \citenamefont {Bulgac}, \citenamefont {Carlson}, \citenamefont {Engel},
    \citenamefont {Fann} \emph {et~al.}}]{Bogner2013}%
    \BibitemOpen
    \bibfield  {author} {\bibinfo {author} {\bibfnamefont {S.}~\bibnamefont
    {Bogner}}, \bibinfo {author} {\bibfnamefont {A.}~\bibnamefont {Bulgac}},
    \bibinfo {author} {\bibfnamefont {J.~A.}\ \bibnamefont {Carlson}}, \bibinfo
    {author} {\bibfnamefont {J.}~\bibnamefont {Engel}}, \bibinfo {author}
    {\bibfnamefont {G.}~\bibnamefont {Fann}},  \emph {et~al.},\ }\href {\doibase
    10.1016/j.cpc.2013.05.020} {\bibfield  {journal} {\bibinfo  {journal}
    {Comput.Phys.Commun.}\ }\textbf {\bibinfo {volume} {184}},\ \bibinfo {pages}
    {2235} (\bibinfo {year} {2013})},\ \Eprint {http://arxiv.org/abs/1304.3713}
    {arXiv:1304.3713 [nucl-th]} \BibitemShut {NoStop}%
  \bibitem [{\citenamefont {Lusk}\ \emph {et~al.}(2010)\citenamefont {Lusk},
    \citenamefont {Pieper},\ and\ \citenamefont {Butler}}]{Lusk2010}%
    \BibitemOpen
    \bibfield  {author} {\bibinfo {author} {\bibfnamefont {E.~L.}\ \bibnamefont
    {Lusk}}, \bibinfo {author} {\bibfnamefont {S.~C.}\ \bibnamefont {Pieper}}, \
    and\ \bibinfo {author} {\bibfnamefont {R.~M.}\ \bibnamefont {Butler}},\
    }\href@noop {} {\bibfield  {journal} {\bibinfo  {journal} {SciDac Rev. B}\
    }\textbf {\bibinfo {volume} {17}},\ \bibinfo {pages} {30} (\bibinfo {year}
    {2010})}\BibitemShut {NoStop}%
  \bibitem [{\citenamefont {Forum}(1993)}]{MPI}%
    \BibitemOpen
    \bibfield  {author} {\bibinfo {author} {\bibfnamefont {T.~M.}\ \bibnamefont
    {Forum}},\ }\href@noop {} {\enquote {\bibinfo {title} {Mpi: A message passing
    interface},}\ } (\bibinfo {year} {1993})\BibitemShut {NoStop}%
  \bibitem [{\citenamefont {Yildirim}\ and\ \citenamefont
    {Ciraci}(2005)}]{Yildirim2005}%
    \BibitemOpen
    \bibfield  {author} {\bibinfo {author} {\bibfnamefont {T.}~\bibnamefont
    {Yildirim}}\ and\ \bibinfo {author} {\bibfnamefont {S.}~\bibnamefont
    {Ciraci}},\ }\href {\doibase 10.1103/PhysRevLett.94.175501} {\bibfield
    {journal} {\bibinfo  {journal} {Phys. Rev. Lett.}\ }\textbf {\bibinfo
    {volume} {94}},\ \bibinfo {pages} {175501} (\bibinfo {year}
    {2005})}\BibitemShut {NoStop}%
  \bibitem [{\citenamefont {Shore}(1990)}]{Shore1990}%
    \BibitemOpen
    \bibfield  {author} {\bibinfo {author} {\bibfnamefont {B.~W.}\ \bibnamefont
    {Shore}},\ }\href@noop {} {\emph {\bibinfo {title} {The Theory of Coherent
    Atomic Excitation, Volume 2, Multilevel Atoms and Incoherence}}}\ (\bibinfo
    {publisher} {Wiley-VCH},\ \bibinfo {year} {1990})\BibitemShut {NoStop}%
  \bibitem [{\citenamefont {Metcalf}\ and\ \citenamefont {van~der
    Straten}(2001)}]{metcalf2001}%
    \BibitemOpen
    \bibfield  {author} {\bibinfo {author} {\bibfnamefont {H.}~\bibnamefont
    {Metcalf}}\ and\ \bibinfo {author} {\bibfnamefont {P.}~\bibnamefont {van~der
    Straten}},\ }\href@noop {} {\emph {\bibinfo {title} {Laser Cooling and
    Trapping}}},\ Graduate Texts in Contemporary Physics\ (\bibinfo  {publisher}
    {Springer New York},\ \bibinfo {year} {2001})\BibitemShut {NoStop}%
  \bibitem [{\citenamefont {Rumble}(2017)}]{CRC}%
    \BibitemOpen
    \bibinfo {editor} {\bibfnamefont {J.}~\bibnamefont {Rumble}},\ ed.,\
    \href@noop {} {\emph {\bibinfo {title} {CRC Handbook of Chemistry and
    Physics, 98th Edition}}}\ (\bibinfo  {publisher} {CRC Press},\ \bibinfo
    {year} {2017})\BibitemShut {NoStop}%
  \bibitem [{\citenamefont {Scoles}(1988)}]{Scoles1988}%
    \BibitemOpen
    \bibinfo {editor} {\bibfnamefont {G.}~\bibnamefont {Scoles}},\ ed.,\
    \href@noop {} {\emph {\bibinfo {title} {{Atomic and Molecular Beam Methods :
    Volume 1 (Atomic \& Molecular Beam Methods)}}}}\ (\bibinfo  {publisher}
    {Oxford University Press, USA},\ \bibinfo {year} {1988})\BibitemShut
    {NoStop}%
  \bibitem [{\citenamefont {Olander}\ and\ \citenamefont
    {Kruger}(1970)}]{Olander1970III}%
    \BibitemOpen
    \bibfield  {author} {\bibinfo {author} {\bibfnamefont {D.~R.}\ \bibnamefont
    {Olander}}\ and\ \bibinfo {author} {\bibfnamefont {V.}~\bibnamefont
    {Kruger}},\ }\href@noop {} {\bibfield  {journal} {\bibinfo  {journal}
    {Journal of Applied Physics}\ }\textbf {\bibinfo {volume} {41}},\ \bibinfo
    {pages} {2769} (\bibinfo {year} {1970})}\BibitemShut {NoStop}%
  \bibitem [{\citenamefont {Olander}\ \emph {et~al.}(1970)\citenamefont
    {Olander}, \citenamefont {Jones},\ and\ \citenamefont
    {Siekhaus}}]{Olander1970IV}%
    \BibitemOpen
    \bibfield  {author} {\bibinfo {author} {\bibfnamefont {D.}~\bibnamefont
    {Olander}}, \bibinfo {author} {\bibfnamefont {R.}~\bibnamefont {Jones}}, \
    and\ \bibinfo {author} {\bibfnamefont {W.}~\bibnamefont {Siekhaus}},\
    }\href@noop {} {\bibfield  {journal} {\bibinfo  {journal} {Journal of Applied
    Physics}\ }\textbf {\bibinfo {volume} {41}},\ \bibinfo {pages} {4388}
    (\bibinfo {year} {1970})}\BibitemShut {NoStop}%
  \bibitem [{\citenamefont {Arpornthip}\ \emph {et~al.}(2012)\citenamefont
    {Arpornthip}, \citenamefont {Sackett},\ and\ \citenamefont
    {Hughes}}]{Arpornthip2012}%
    \BibitemOpen
    \bibfield  {author} {\bibinfo {author} {\bibfnamefont {T.}~\bibnamefont
    {Arpornthip}}, \bibinfo {author} {\bibfnamefont {C.~A.}\ \bibnamefont
    {Sackett}}, \ and\ \bibinfo {author} {\bibfnamefont {K.~J.}\ \bibnamefont
    {Hughes}},\ }\href {\doibase 10.1103/PhysRevA.85.033420} {\bibfield
    {journal} {\bibinfo  {journal} {Phys. Rev. A}\ }\textbf {\bibinfo {volume}
    {85}},\ \bibinfo {pages} {033420} (\bibinfo {year} {2012})}\BibitemShut
    {NoStop}%
  \bibitem [{\citenamefont {Teodoro}\ \emph {et~al.}(2015)\citenamefont
    {Teodoro}, \citenamefont {Haiduke}, \citenamefont {Dammalapati},
    \citenamefont {Knoop},\ and\ \citenamefont {Visscher}}]{Teodoro2015}%
    \BibitemOpen
    \bibfield  {author} {\bibinfo {author} {\bibfnamefont {T.~Q.}\ \bibnamefont
    {Teodoro}}, \bibinfo {author} {\bibfnamefont {R.~L.~A.}\ \bibnamefont
    {Haiduke}}, \bibinfo {author} {\bibfnamefont {U.}~\bibnamefont
    {Dammalapati}}, \bibinfo {author} {\bibfnamefont {S.}~\bibnamefont {Knoop}},
    \ and\ \bibinfo {author} {\bibfnamefont {L.}~\bibnamefont {Visscher}},\
    }\href {\doibase 10.1063/1.4929348} {\bibfield  {journal} {\bibinfo
    {journal} {The Journal of Chemical Physics}\ }\textbf {\bibinfo {volume}
    {143}},\ \bibinfo {pages} {084307} (\bibinfo {year} {2015})}\BibitemShut
    {NoStop}%
  \bibitem [{\citenamefont {Porsev}\ and\ \citenamefont
    {Derevianko}(2006)}]{Porsev2006}%
    \BibitemOpen
    \bibfield  {author} {\bibinfo {author} {\bibfnamefont {S.}~\bibnamefont
    {Porsev}}\ and\ \bibinfo {author} {\bibfnamefont {A.}~\bibnamefont
    {Derevianko}},\ }\href {\doibase 10.1134/S1063776106020014} {\bibfield
    {journal} {\bibinfo  {journal} {J. Exp. Theor. Phys.}\ }\textbf {\bibinfo
    {volume} {102}},\ \bibinfo {pages} {195} (\bibinfo {year}
    {2006})}\BibitemShut {NoStop}%
  \bibitem [{\citenamefont {Silvera}\ and\ \citenamefont
    {Goldman}(1978)}]{Silvera1978}%
    \BibitemOpen
    \bibfield  {author} {\bibinfo {author} {\bibfnamefont {I.~F.}\ \bibnamefont
    {Silvera}}\ and\ \bibinfo {author} {\bibfnamefont {V.~V.}\ \bibnamefont
    {Goldman}},\ }\href {\doibase 10.1063/1.437103} {\bibfield  {journal}
    {\bibinfo  {journal} {The Journal of Chemical Physics}\ }\textbf {\bibinfo
    {volume} {69}},\ \bibinfo {pages} {4209} (\bibinfo {year}
    {1978})}\BibitemShut {NoStop}%
  \bibitem [{\citenamefont {Bissonnette}\ \emph {et~al.}(1996)\citenamefont
    {Bissonnette}, \citenamefont {Chuaqui}, \citenamefont {Crowell},
    \citenamefont {Roy}, \citenamefont {Wheatley},\ and\ \citenamefont
    {Meath}}]{Bissonnette1996}%
    \BibitemOpen
    \bibfield  {author} {\bibinfo {author} {\bibfnamefont {C.}~\bibnamefont
    {Bissonnette}}, \bibinfo {author} {\bibfnamefont {C.~E.}\ \bibnamefont
    {Chuaqui}}, \bibinfo {author} {\bibfnamefont {K.~G.}\ \bibnamefont
    {Crowell}}, \bibinfo {author} {\bibfnamefont {R.~J.~L.}\ \bibnamefont {Roy}},
    \bibinfo {author} {\bibfnamefont {R.~J.}\ \bibnamefont {Wheatley}}, \ and\
    \bibinfo {author} {\bibfnamefont {W.~J.}\ \bibnamefont {Meath}},\ }\href
    {\doibase 10.1063/1.472127} {\bibfield  {journal} {\bibinfo  {journal} {The
    Journal of Chemical Physics}\ }\textbf {\bibinfo {volume} {105}},\ \bibinfo
    {pages} {2639} (\bibinfo {year} {1996})}\BibitemShut {NoStop}%
  \bibitem [{\citenamefont {Aziz}\ and\ \citenamefont {Chen}(1977)}]{Aziz1977}%
    \BibitemOpen
    \bibfield  {author} {\bibinfo {author} {\bibfnamefont {R.~A.}\ \bibnamefont
    {Aziz}}\ and\ \bibinfo {author} {\bibfnamefont {H.~H.}\ \bibnamefont
    {Chen}},\ }\href {\doibase 10.1063/1.434827} {\bibfield  {journal} {\bibinfo
    {journal} {The Journal of Chemical Physics}\ }\textbf {\bibinfo {volume}
    {67}},\ \bibinfo {pages} {5719} (\bibinfo {year} {1977})}\BibitemShut
    {NoStop}%
  \end{thebibliography}
%

\appendix

\section{Calculation of photon absorption rates}\label{app:rates}
We decompose the laser field into an array of circularly and linearly 
polarized beams, all of which compete for the excitation of the atom.  The 
beams are described by a reference point $\va{x}$, a propagation direction $\va
{p}$, transverse elliptical size ($r_1$, $r_2$), Gaussian intensity profile 
($\sigma_1$, $\sigma_2$), divergence angle ($d_1$, $d_2$), polarization 
helicity ($h=\pm 1$ for circularly polarized and $h=0$ for linearly polarized),
 saturation intensity $\beta = I/I_{sat}$ and detuning $\Gamma = \delta/
\gamma$ where $\delta$ is the bare laser detuning and $\gamma=1/\tau$ is the 
natural linewidth with $\tau$ the upper state lifetime.

We determine the intensity $\beta_i$ and photon direction $\va{p}_i$ of each 
beam (indicated by the subscript $i$) at the location of the atom.  The Zeeman 
shift of the upper (lower) state in natural linewidths per unit $m_F$ is~\cite
{Shore1990}:
\begin{equation}\label{eq:zeeman}
    Z_{u(l)}=\frac{\mu_Bg_{u(l)}}{\hbar}B\tau
\end{equation}
where $\mu_B$ is the Bohr magneton and $\hbar$ the reduced Planck constant, $g_
{u(l)}$ is the Land\`e g-factor of the upper (lower) state, $\tau$ is the 
natural lifetime of the upper state, and $B\equiv\qty\big|\va{B}|$ is the 
magnetic field amplitude.

The Doppler shift $D_i$ in natural linewidths is given by:
\begin{equation}\label{eq:doppler}
    D_u = -2\pi\frac{\tau}{\lambda}\va{v}\vdot\va{p}_i
\end{equation}
where $\vb{v}$ is the velocity of the atom and $\lambda$ is the wavelength of 
the photon.

The angular distribution for the transition probability~\cite{Shore1990} $W_{i,
\Delta m}$ for circularly polarized light ($h=\pm 1$) is:
\begin{align}\label{eq:deltam_circular}
    W_{i,1} &=  \tfrac{1}{4}\qty(1 +h \cos\qty(\theta_i))^2\nonumber \\
    W_{i,0} &= \tfrac{1}{2}\sin^2\qty(\theta_i)\\
    W_{i,-1} &= \tfrac{1}{4}\qty(1 - h \cos\qty(\theta_i))^2\nonumber
\end{align}
and for linearly polarized light ($h=0$) by:
\begin{align}\label{eq:deltam_linear}
    W_{i,1} &= \tfrac{1}{2}\cos^2\qty(\theta_i)\nonumber \\
    W_{i,0} &= \sin^2\qty(\theta_i)\\
    W_{i,-1} &= \tfrac{1}{2}\cos^2\qty(\theta_i)\nonumber
\end{align}
where $\theta_i$ is the angle between $\va{B}$ and $\va{p}_i$, and $\Delta m = 
0, \pm 1$ is the change in $m_F$ for the transition.  We treat all linear 
polarization as vertical since there is no preferred direction orthogonal to 
$\va{p}_i$ in the frame of the laser beam.  We sum over the contributions of 
$W_{i,\Delta m}$ to find the effective intensity of each laser $i$:
\begin{align}\label{eq:eff_intensity}
    \beta_i ' &= \sum_{\Delta m = -1}^1 \beta_{i,\Delta m} '\nonumber\\
    &= \sum_{\Delta m = -1}^1 \frac{\beta_i W_{i,\Delta m}}{1+4\qty[\Gamma_i + 
        D_i + m_F Z_l - \qty(m_F+\Delta m)Z_u]^2}
\end{align}

We model a non-zero laser linewidth by re-defining the saturation parameter 
$\beta$ as
\begin{equation}\label{eq:satparam}
    \beta = \beta_0 \int_{-\infty}^{\infty}
        \frac{L(x)\dd{x}}{1+4\qty(x+\Gamma)^2}
\end{equation}
where $\beta_0 = I/I_{\text{sat}}$ is the on-resonance saturation parameter, $L
(x)$ is a normalized Lorentzian function, and $\Gamma=\delta/\gamma_{\text{Ra}}
$ is the detuning.  Evaluating the integral we find
\begin{equation}\label{eq:betafrac}
    \frac{\beta}{\beta_0} = \frac{1+w}{4\Gamma^2+(1+w)^2}
\end{equation}
where $w$ is the full-width at half-maximum of the Lorentzian.  In the limit 
$w \rightarrow 0$, we recover the familiar definition of the saturation 
parameter~\cite{metcalf2001}.  This effectively broadens the absorption 
spectrum, leading to increased excitation probability at large detuning, but 
reduced excitation probability close to resonance.

The total absorption rate ($R_T$) and fractional absorption rate ($R_{i,\Delta 
m}$) for each beam $i$ are:
\begin{align}\label{eq:absorption_rates}
    R_T &= \frac{\beta_T '}{\qty(2+\beta_T ')\tau}\nonumber\\
    R_{i,\Delta m} &= R_T\qty(\frac{\beta_{i,\Delta m} '}{\beta_T '})
\end{align}
where $\beta_T ' = \sum_i \beta_i '$ is the total intensity.  We emphasize 
that this is the photon absorption rate for the ground state of the atom; once 
the atom is excited, we set the absorption rate to zero and the emission rate 
to $\gamma$, ignoring stimulated emission.

\section{Effusive oven velocity and angular distributions}\label{app:distros}
For an an effusive oven with a Knudsen number $K_n = \lambda_{\text{atom}}/L_
{\text{noz}}$, we calculate the mean free path of the atom:
\begin{equation}\label{eq:mfp}
    \lambda_{\text{atom}}=\frac{k_b T}{\sqrt{2}\pi d^2 P}
\end{equation}
where $k_B$ is the Boltzmann constant, $T$ is the temperature, $d$ is the 
interaction distance between two atoms (taken as the sum of the van der Waals 
radii), and $P$ is the saturated vapor pressure at a temperature $T$, 
calculated as~\cite{CRC}:
\begin{equation}\label{eq:svp}
    \log_{10}\qty(\frac{P}{X})=A+\frac{B}{T}+C\log_{10}T+\frac{D}{T^3}
\end{equation}
where $X$ is the pressure per atmosphere in the desired units, and $A$, $B$, 
$C$, and $D$ are experimentally obtained coefficients available in Ref.~\cite
{CRC}.

We determine the velocity distribution of the atoms in the simulation by the 
temperature $T$ of the last wall in the oven or nozzle that an atom collides 
with as~\cite{Scoles1988,Olander1970III,Olander1970IV}:
\begin{equation}\label{eq:MBkn}
    f(u) = A\qty[f_{\text{beam}}(u)\mathcal{P}\qty(K_n,\psi(u))]
\end{equation}
where $u$ is the reduced velocity $v_z/\tilde v$ ($v_z$ is the longitudinal 
component of the velocity) and $\tilde v \equiv \sqrt{2 k_B T / M}$ is the 
most probable velocity where $M$ the mass of the atom.  $A$ is a normalization 
constant satisfying $\int_0^{\infty}f(u)\ du = 1$.  The unperturbed Boltzmann 
distribution for an atomic beam is:
\begin{equation}\label{eq:MBspeed}
    f_{\text{beam}}(u)=\frac{2u^3}{\tilde v}\exp\qty(-u^2)
\end{equation}
and the Knudsen number perturbation is:
\begin{equation}\label{eq:pert}
    \mathcal{P}\qty(K_n,\psi(u)) = \frac{\sqrt{\pi}}{2}
        \frac{\erf\sqrt{\psi(u)/2K_n}}{\sqrt{\psi(u)/2K_n}}
\end{equation}
with the function $\psi(u)$ defined as:
\begin{equation}\label{eq:psiu}
    \psi(u) = \frac{ue^{-u^2}+\qty(\sqrt{\pi}/2)\qty(1+2u^2)
        \erf(u)}{\sqrt{2\pi}u^2}
\end{equation}
We include both the molecular and intermediate flow regime distributions in 
the simulation and reproduce the formulas from Refs.~\cite{Scoles1988,
Olander1970III} below.  The molecular flow angular distribution depends solely 
on the oven nozzle's geometry, and for polar angles $\theta$ from the axis of 
the atomic beam, the distribution of angles originating from the oven 
($\chi_\mathrm{AR}\ge\tan\theta$ where $\chi_\mathrm{AR} = d_{\text{noz}}/L_
{\text{noz}}$) is:
\begin{align}\label{eq:jml}
    j_{\mathcal{M}}(\theta) = \zeta_0\cos\theta&+\frac{2}{\pi}
        \qty[(1-\zeta_0)R(q)]\cos\theta \nonumber \\
    &+\frac{4}{3\pi q}\qty[1-\qty(1-q^2)^{3/2}]\qty[\zeta_1-\zeta_0]\cos\theta
\end{align}
and for angles originating from the nozzle wall ($\chi_\mathrm{AR}
\le\tan\theta$):
\begin{align}\label{eq:jmh}
    j_{\mathcal{M}}(\theta) = \zeta_0\cos\theta 
        + \frac{4}{3\pi q}(\zeta_1-\zeta_0)\cos\theta
\end{align}
where the following definitions are used:
\begin{align}
    q &= \tan\theta/\chi_\mathrm{AR}\label{eq:jq}\\
    R(q) &= \cos^{-1}(q) - q\sqrt{1-q^2}\label{eq:jR}
\end{align}
and the collision rates with the nozzle walls at the exit and entrance, 
$\zeta_0$ and $\zeta_1$, respectively, are two dimensionless parameters 
defined as:
\begin{align}\label{eq:z0z1}
    \zeta_0 &= \zeta_1-\frac{\qty(1+2/\chi_\mathrm{AR})
        \sqrt{1+1/\chi_\mathrm{AR}^2}-\qty(1+2/\chi_\mathrm{AR}^2)}
        {\sqrt{1+1/\chi_\mathrm{AR}^2}+1}\nonumber\\
    \zeta_1 &= \qty[1+\chi_\mathrm{AR}/(2+\chi_\mathrm{AR}^2)]^{-1}
\end{align}

The intermediate flow regime angular distribution accounts for both the 
nozzle's geometry and interatomic collisions in the nozzle.  This distribution 
depends on the Knudsen number as well as the number density of the atoms at 
the exit and entrance of the nozzle, characterized by the two dimensionless 
parameters $\xi_0$ and $\xi_1$, which are conventionally defined as $\xi_0 
\equiv \zeta_0$ and $\xi_1 \equiv \zeta_1$.  The distribution for $\chi_\mathrm
{AR}\ge\tan\theta$ is:
\begin{align}\label{eq:jil}
    j_{\mathcal{I}}(\theta)&=\xi_0\cos\theta\qty[1+\frac{2}{\sqrt{\pi}}
        \frac{e^{\delta^{{'}2}}}{\delta^{'}}S(q)] \nonumber \\
    &+\frac{\xi_0\cos\theta}{\sqrt{\pi}}\frac{e^{\delta^{{'}2}}}{\delta^{'}}
        \qty[R(q)\qty(\erf\xi^{'}-\erf\delta^{'}+F(\xi_0,\xi_1,\delta^{'}))]
\end{align}
and for $\chi_\mathrm{AR}\le\tan\theta$:
\begin{align}\label{eq:jih}
    j_{\mathcal{I}}(\theta)=\xi_0\cos\theta\qty[1+\frac{2}{\sqrt{\pi}}
        \frac{e^{\delta^{{'}2}}}{\delta^{'}}S(1)]
\end{align}
and at $\theta=0$:
\begin{align}\label{eq:ji0}
    j_{\mathcal{I}}(0)&=\xi_0+\frac{\sqrt{\pi}}{2}\xi_0
        \frac{e^{\delta^2}}{\delta}\qty\bigg(\erf\xi-\erf\delta) \nonumber \\
    &+\qty(\frac{1-\xi_1}{\xi_0})e^{-\qty(\xi^2-\delta^2)}
\end{align}
where the following definitions are used:
\begin{equation}\label{eq:Jdelta}
    \delta = \sqrt{\frac{\xi_0^2}{2K_{n}\qty(\xi_1-\xi_0)}}\qq{,}\xi 
        = \qty(\frac{\xi_1}{\xi_0})\delta
\end{equation}
\begin{equation}\label{eq:Jdeltap}
    \delta^{'} = \sqrt{\frac{\delta^2}{\cos\theta}}\qq{,}\xi^{'} 
        = \qty(\frac{\xi_1}{\xi_0})\delta^{'}
\end{equation}
\begin{equation}\label{eq:JIF}
    F(\xi_0,\xi_1,\delta^{'}) = \frac{2}{\sqrt{\pi}}\xi^{'}
        \qty(\frac{1}{\xi_1}-1)e^{-\xi^{'2}}
\end{equation}
\begin{align}\label{eq:JS}
    S(q) = \int^q_0 &\dd z \sqrt{1-z^2}\nonumber \\
    &\times\qty(\erf\qty{\delta^{'}\qty[1+\frac{z}{q}
        \qty(\frac{\xi_1}{\xi_0}-1)]}-\erf\delta^{'})
\end{align}

\section{Estimate of long-range van der Waals $C_6$ coefficients}\label{app:c6coeff}
Residual gas collisions alter an atom's trajectory and can eject atoms from 
the MOT.  To assess the impact of these effects on the optical trapping 
efficiency, we included a non-zero residual gas pressure in the simulation.  
For relevant residual gases in the \Ra~optical trap, e.g.~H$_2$ and Ar, no 
data exists for the $C_6$ coefficient with Ra. Instead, we estimate the 
coefficients with the London dispersion formula:
\begin{equation}\label{eq:london}
    C_6 = \frac{3}{2}\qty(\frac{U_1 U_2}{U_1 + U_2})\alpha_1\alpha_2
\end{equation}
where $\alpha_i$ is the polarizability (in $\si{\angstrom\cubed}$) and $U_i$ a 
relevant energy scale for each constituent.  For tightly bound species such as 
molecules and noble gases, we estimate $U_i$ as the ionization energy, whereas 
for alkaline earths, such as Ra and Ba, we estimate $U_i$ as the energy of the 
$^1 S_0\, \rightarrow\, ^1 P_1$ transition, as this state dominates the ground 
state polarizability.  We then calculate a table of $C_6$ coefficients from 
this estimate, which agree to within $\SI{10}{\percent}$ with accepted values, 
see Tab.~\ref{tbl:c6}.  Since the total scattering cross section scales as 
$\qty(C_6)^{1/3}$~\cite{Arpornthip2012}, calculations of collisional processes 
are only weakly sensitive to small errors in this coefficient.  We also 
require a bond length to generate a Lennard-Jones potential for each pair of 
objects.  In the case of the \Ra~simulation, we estimate this by adding the 
atomic radius of Ra ($\SI{2.15}{\angstrom}$) to the effective radius of the 
other object.  While not precise, in these simulations the final results are 
insensitive to the exact details of the short-range interaction since such a 
low impact parameter collision is certain to alter the trajectory and/or eject 
the \Ra~atom from the MOT.

\begin{table}[ht]
\centering
\bgroup
\def\arraystretch{1.1}
\begin{tabular}{|c|c|c|c|c|}
\hline & Ra & Ba & H$_2$ & Ar \\
\hline
$U(\si{\eV})$ & 2.57 & 2.53 & 15.43 & 15.76 \\
$\alpha (\si{\angstrom}^3)$ & 37.9 & 39.7 & 0.8042 & 1.63\\
\hline
Ra & 2770 & 2870 & 101 & 205\\
& {\em 3042~\cite{Teodoro2015}} & & & \\
\hline
Ba & & 2980 & 104 & 211 \\
& & {\em 3110~\cite{Porsev2006}} & & \\
\hline
H$_2$ & & & 7.48 & 15.3 \\
& & & {\em 7.29~\cite{Silvera1978}} &{\em 16.6~\cite{Bissonnette1996}} \\
\hline
Ar & & & & 31.3 \\
& & & & {\em 34.8~\cite{Aziz1977}} \\
\hline
\end{tabular}
\egroup
\caption{\label{tbl:c6}
Table of $C_6$ coefficients calculated using the London dispersion formula.  
All entries are in $\si{\eV}\, \si{\angstrom}^6$.  Accepted values listed in 
italics and are in agreement with our estimated values to within $\SI{10}
{\percent}$.  The values for H$_2$ are for its vibrational ground state.}
\end{table}

\end{document}